\documentclass[journal]{IEEEtran}

\usepackage[cmex10]{amsmath}
\usepackage{amssymb}
\usepackage{cite}
\usepackage{amsmath}
\usepackage{amssymb}
\usepackage{wasysym}
\usepackage{booktabs}
\usepackage{paralist}
\usepackage{balance}
\usepackage{subfigure}
\usepackage{multicol}
\usepackage{graphicx}
\usepackage{epsfig}
\usepackage{epstopdf}
\usepackage[ruled,linesnumbered]{algorithm2e}
\usepackage{bm}
\usepackage{amsmath, amssymb,amsthm}
\usepackage{upgreek}
\DeclareMathOperator*{\argmax}{argmax}

\newtheorem{proposition}{Proposition}
\newtheorem{lemma}{Lemma}

\ifCLASSINFOpdf
\else
\fi
\begin{document}
%
\title{Resource Allocation in Dynamic TDD Heterogeneous Networks under Mixed Traffic}
%
%
%

\author{Qiang~Fan,
        Hancheng~Lu,~\IEEEmembership{Member,~IEEE,}
        Peilin~Hong,~\IEEEmembership{Senior~Member,~IEEE,} 
        and Chang Wen Chen,~\IEEEmembership{Fellow,~IEEE}
\IEEEcompsocitemizethanks{\IEEEcompsocthanksitem Qiang Fan, Hancheng Lu and Peilin Hong are with the Information Network Lab of EEIS Department, University of Science and Technology of China, Hefei 230027, China (Email: fanq@mail.ustc.edu.cn, \{hclu, plhong\}@ustc.edu.cn).
\IEEEcompsocthanksitem Chang Wen Chen is with the State University of New York, Buffalo, NY 14260 USA (Email: Chencw@buffalo.edu).

This work is supported in parts by the National Science Foundation of China (No.61390513, No.91538203), the National High Technology Research and Development Program of China (863 Program) (No.2014AA01A706).}
}

\maketitle

\begin{abstract}
Recently, Dynamic Time Division Duplex (TDD) has been proposed to handle the asymmetry of traffic demand between DownLink (DL) and UpLink (UL) in Heterogeneous Networks (HetNets). However, for mixed traffic consisting of best effort traffic and soft Quality of Service (QoS) traffic, the resource allocation problem has not been adequately studied in Dynamic TDD HetNets. In this paper, we focus on such problem in a two-tier HetNet with co-channel deployment of one Macro cell Base Station (MBS) and multiple Small cell Base Stations (SBSs) in hotspots. Different from existing work, we introduce low power almost blank subframes to alleviate MBS-to-SBS interference which is inherent in TDD operation.
To tackle the resource allocation problem, we propose a two-step strategy. First, from the view point of base stations, we propose a transmission protocol and perform time resource allocation by formulating and solving a network capacity maximization problem under DL/UL traffic demands. Second, from the view point of User Equipments (UEs), we formulate their resource allocation as a Network Utility Maximization (NUM) problem. An efficient iterative algorithm is proposed to solve the NUM problem. Simulations show the advantage of the proposed algorithm in terms of network throughput and UE QoS satisfaction level.

\end{abstract}


%
\IEEEpeerreviewmaketitle

\section{Introduction}
%
%
%
%
\IEEEPARstart{I}{N} recent years, the popularization of smart mobile devices and the development of various multimedia services have led to an exponential surge in mobile traffic volume. To alleviate traffic pressure of a traditional cellular cell (i.e., macro cell), small cells are proposed and deployed in hotspots to offload traffic to/from the Macro cell Base Station (MBS). Such two different types of cells form a Heterogeneous Network (HetNet) \cite{N.Wang}. HetNet increases the network capacity by cell densification with spatial frequency reuse. It has been proved to be a potential solution to meet the increasing traffic demands in future fifth Generation (5G) cellular networks \cite{J.G.Andrews}.

Within a HetNet, Use Equipments (UEs) with mixed traffic usually have different Quality of Service (QoS) requirements. In this paper, mixed traffic consists of two kinds of traffic, i.e., Best Effort (BE) traffic and soft QoS traffic \cite{L.Chen}. Without loss of generality, consider the QoS requirements can be expressed in terms of service rates. In this case, the diversity of the QoS requirements results in traffic demand asymmetry between DownLink (DL) and UpLink (UL). Moreover, traffic demand asymmetry is time-varying, especially in small cells deployed in hotspots \cite{M.Ding_1}. In such scenarios, Time Division Duplex (TDD) is selected over Frequency Division Duplex (FDD) for its inherent ability to handle variation of DL and UL traffic demands \cite{C.Yoon}\cite{N.Zorba}. Compared with FDD where frequency bands for DL and UL transmissions are statically assigned, TDD can easily allocate the fraction of time dedicated to UL and DL transmissions on a per-frame basis according to the traffic demands, which is referred to as dynamic TDD \cite{H.Sun}\cite{A.K.Gupta}.

Dynamic TDD has been considered as a promising technology in future 5G cellular networks where HetNets with mixed traffic will be widely deployed \cite{B.Yu}\cite{Y.Zhong}. However, it also faces two major challenges. Consider a two-tier HetNet with co-channel deployment of one macro cell and multiple small cells in hotspots. First, in the dynamic TDD HetNet, all of the DL and UL transmissions operate within the same spectrum. Due to asynchronous configuration among neighboring cells, cross-link interference, i.e., DL-to-UL interference and UL-to-DL interference \cite{Z.Shen}, will be brought in. In practical deployment, MBS employs much more power for transmissions than Small cell Base Stations (SBSs) in hotspots. Thus, DL-to-UL interference produced by DL transmissions from MBS and UL transmissions to SBS, which can be referred to as MBS-to-SBS interference, is severe and leads to a significant performance degradation in UL transmissions to SBS. Second, traffic demands in small cells deployed in hotspots usually suffer much more fluctuations than that in the macro cell \cite{M.Ding_3}. When dynamic TDD is simultaneously operated at MBS and SBSs, SBSs will require more flexibility in the configuration of UL and DL transmission resources than the MBS.

To overcome the challenges we have just discussed in the dynamic TDD HetNet, we propose a Low Power Almost Blank Subframes (LP-ABS) based resource allocation mechanism. Different from existing work, we consider the scenario under mixed traffic. With LP-ABS, MBS will reduce its transmission power when cross-link interference between MBS and SBSs occurs. Consequently, MBS-to-SBS interference is alleviated. Moreover, the application of LP-ABS brings additional flexibility to combat traffic fluctuations in small cells by employing power allocation at MBS and configuration of UL and DL subframes at SBSs. In previous work, Almost Blank Subframes (ABS) \cite{S.Deb} have been exploited to mitigate MBS-to-SBS interference \cite{M.Ding_3,M.Kamel_1,M.Kamel_2,H.Ji}. ABS are subframes that contain only common reference signals and the most important cell-specific broadcast information. MBS schedules ABS in the DL direction when the interfered UEs in small cells perform UL transmissions to SBSs. Thus, severe MBS-to-SBS interference can be suppressed. However, ABS based schemes sacrifice precious transmission opportunities at MBS. We observe that UEs with mixed traffic have different QoS requirements in terms of service rates. In practical wireless environments, UEs can endure certain interference as long as their requirements on service rates can be satisfied \cite{W.Lu}. This observation inspires us to apply LP-ABS, instead of ABS, at MBS. With LP-ABS, rather than blanking, MBS adopts a proper power reduction in DL transmissions subject to the diverse QoS requirements from UEs. By doing so, we can further improve the spectral efficiency in the dynamic TDD HetNet while alleviating MBS-to-SBS interference. Furthermore, during the LP-ABS period, SBSs can adjust the configuration of UL and DL transmission resources to meet the varying DL/UL traffic demands. At the same time, MBS can also adjust transmission power in the DL direction based on the traffic conditions in HetNet. These two adjustment procedures are coupled and can be jointly optimized to exploit the diversity of the QoS requirements.

The introduction of LP-ABS shall make the resource allocation problem in the dynamic TDD HetNet under mixed traffic more complicated. Two kinds of resources, i.e., time resources, in forms of subframes, and power, need to be allocated to meet the DL/UL traffic demands. However, allocation of subframes and power that are coupled with each other should be jointly optimized. In particular, the coupling relation in resource allocation is reflected in the following key problems. First, how many subframes should be allocated for LP-ABS operation? Second, how much power should be used when MBS performs LP-ABS. Third, under mixed traffic, how many subframes should be allocated to each UE to meet its DL/UL traffic demands? To tackle these problems, we propose a two-step strategy. The basic idea is to perform resource allocation at coarse granularity level (i.e., BS-level) first and then in fine granularity level (i.e., UE-level). By doing so, the complexity of the resource allocation problem can be adequately reduced.

The main contributions of this research can are summarized as follows.

1) To the best of our knowledge, we are the first to investigate the resource allocation problem in the dynamic TDD HetNet under mixed traffic. Moreover, different from existing work, we introduce LP-ABS to alleviate MBS-to-SBS interference. The application of LP-ABS is able to improve the spectral efficiency and bring additional flexibility to combat traffic fluctuations in small cells.

2) For the dynamic TDD HetNet, we propose an effective transmission protocol for the macro cell and small cells where each transmission cycle is divided into a normal DL/UL period and a LP-ABS period. In order to determine the duration of these two periods, we perform time resource allocation at BS-level by formulating and solving a network capacity maximization problem under DL/UL traffic demands.

3) Considering the nature of mixed traffic, we introduce utility functions to represent the satisfaction levels of traffic flows at UEs. Then, based on the results from time resource allocation at BS-level, we formulate the resource allocation problem at UE-level as a Network Utility Maximization (NUM) problem, where the set of subframes allocated to each UE as well as the power used by MBS during the LP-ABS period are properly incorporated.

4) We propose an efficient iterative algorithm to solve the NUM problem. The key idea is to decompose the target problem into multiple cell utility maximization problems under given LP-ABS power and then find the optimal solution by iteratively allocating the LP-ABS power. Each cell utility maximization problem aims at maximizing the sum of utilities for all UEs within a cell. We prove the necessary condition of the optimal solution and solve it independently at each BS. Furthermore, we also prove that the complexity of the proposed algorithm is polynomial.

Numerical results have been obtained to validate the proposed analytical and algorithmic work. Comparing with existing schemes, the proposed algorithm can achieve more network throughput as well as better QoS satisfication levels for UEs.

The rest of the paper is organized as follows: Section II introduces the reviews of the related work in the literature. Section III presents the system model. The problem formulation and extensive analysis are illustrated in Section IV. Besides, a low complexity resource allocation algorithm is designed and given in Section IV. Numerical simulation results are presented and discussed in Section V. Finally, we conclude this paper in Section VI.

\section{Related Work}
Dynamic TDD has been considered as a promising technology in 5G cellular networks where HetNets with mixed traffic will be deployed.
How to alleviate MBS-to-SBS interference and how to allocate resource under mixed traffic are two challenging problems in dynamic TDD HetNets. These two problems are related in that the results of resource allocation will affect the MBS-to-SBS interference level and pattern. There exists some work on the former. Several  interference mitigation mechanisms have been proposed. However, the latter has not been well studied so far.

Synchronous configuration among neighbour cells has been widely adopted as an interference mitigation mechanism for TDD HetNets \cite{Z.Shen,P.Tarasak,D.Liu}.  The widely deployed TD-LTE \cite{Z.Shen} systems typically adopted synchronous configuration across the entire network. In \cite{P.Tarasak}, a cell selection scheme based on DL-UL capacity was proposed for TDD HetNets, where synchronization was assumed for all transmissions in both directions. Authors in \cite{D.Liu} proposed an optimal backhaul-aware joint UL and DL user association for TDD HetNets, where synchronous operation was adopted to eliminate BS-to-BS and UE-to-UE interference. 
Besides, in \cite{H.Sun}, MBSs and SBSs were operated on orthogonal frequency bands. The cross-link interference was eliminated at the expense of additional scarce spectrum.

In \cite{H.Ji,M.Ding_3,M.Kamel_1,M.Kamel_2}, ABS based interference mitigation mechanisms have been introduced for dynamic TDD HetNets. In these mechanisms, MBSs blank some subframes as ABS to avoid severe interference to small cells. In \cite{H.Ji}, MBSs and SBSs configured synchronous DL and UL transmissions on non-ABS, and SBSs applied dynamic TDD on ABS. Similarly, in \cite{M.Ding_3,M.Kamel_1,M.Kamel_2}, MBSs and SBSs configured synchronous DL transmissions on non-ABS. However, SBSs applied dynamic TDD on not only ABS, but also subframes where MBSs configured UL transmissions.

As discussed above, synchronous configuration cannot adapt to DL and UL traffic asymmetry in HetNets. Besides, the ABS based mechanisms sacrifice
transmission opportunities at MBS, which will cause a waste of spectral resource. Furthermore, resource allocation, especially under mixed traffic, has not been well investigated in dynamic TDD HetNets. In fact, resource allocation and interference alleviation are related and they should be jointly considered.
\begin{figure}[t]
  \centering
  \includegraphics[width=0.38\textwidth]{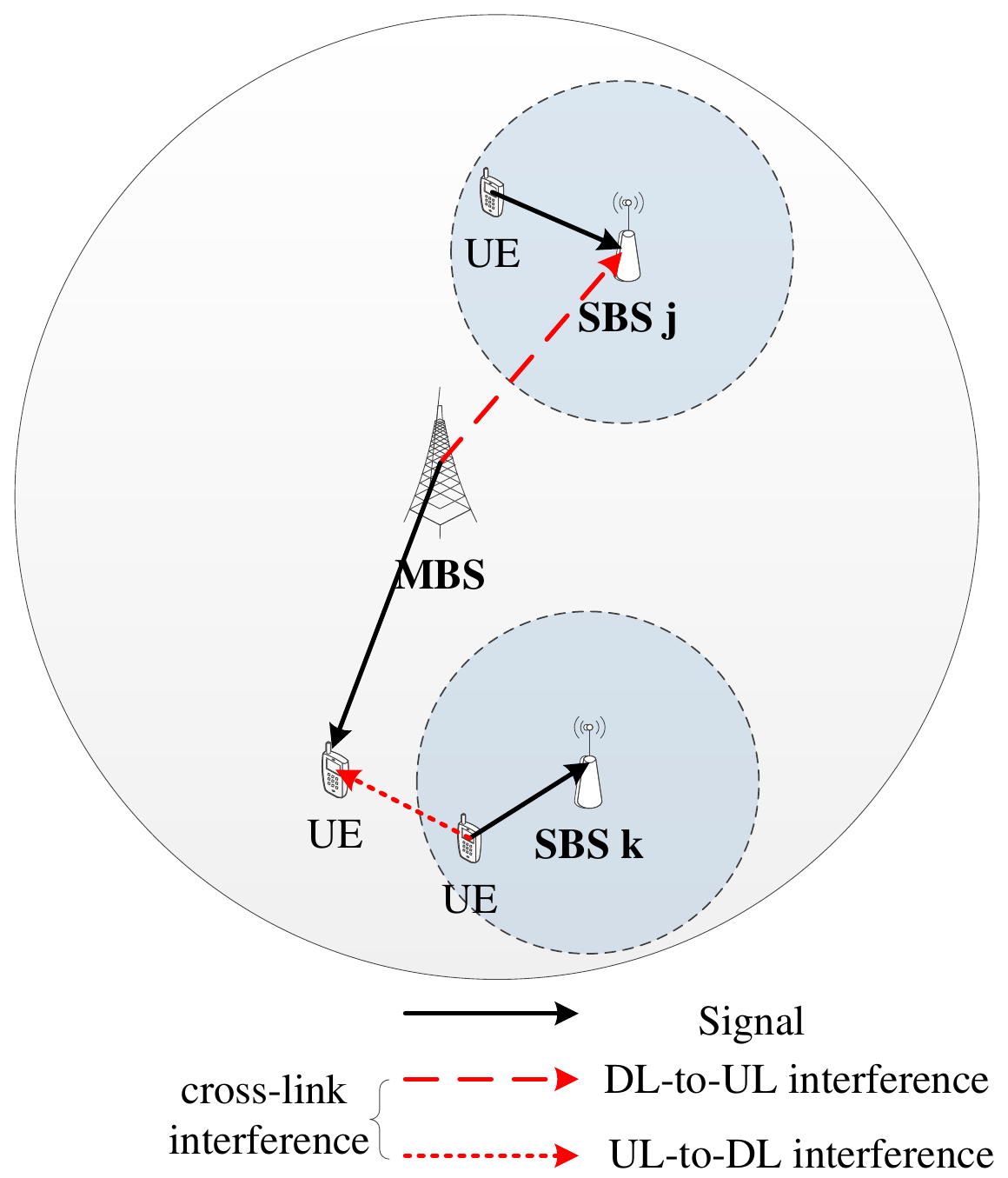}\\
  \caption{A dynamic TDD HetNet with a single MBS and several SBSs. The most serious cross-link interference is the MBS DL-to-SBS UL interference. }\label{network_arch}
\end{figure}

\section{SYSTEM MODEL}
In this paper, the dynamic TDD HetNet shown in Fig. \ref{network_arch} is considered, where a single MBS located in the center of the cell and $N_s$ SBSs are deployed within the range of the macro cell. We denote the set of all BSs by $\mathcal{S}=\{0,1,\ldots,N_s\}$, where index $0$ is introduced for the MBS. Biased maximal received power user association \cite{Y.Lin} is adopted. Let $\mathcal{M}$ and $\mathcal{M}_{j}$ denote the set of all UEs in the HetNet and the set of UEs associated to BS $j$, respectively. Besides, $S_i(\in \mathcal{S})$ represents the associated BS of UE $i(\in\mathcal{M})$. All BSs share the same spectrum.

Throughout this paper, the superscripts `n', `l', `d' and `u' denote normal subframes, LP-ABS, DL transmissions and UL transmissions, respectively.

\begin{figure}[t]
  \centering
  \includegraphics[width=0.35\textwidth]{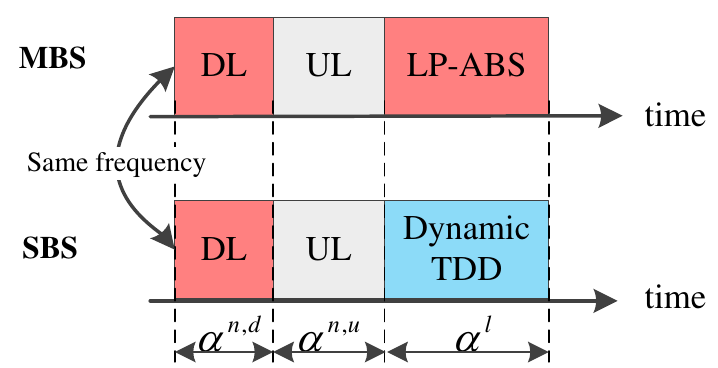}\\
  \caption{Transmission protocol of the proposed LP-ABS configuration}\label{frame_struc}
\end{figure}

\subsection{LP-ABS Mechanism}
In this paper, we propose a LP-ABS interference mitigation mechanism. The transmission protocol is sketched in Fig. \ref{frame_struc}. In each transmission cycle, the MBS divides the subframes into two groups, i.e., \emph{normal subframes} and \emph{LP-ABS}. The MBS configures DL and UL transmissions on normal subframes{\footnote{The normal subframes configured for DL and UL transmissions are referred as normal DL subframes and normal UL subframes, respectively.}}, and DL transmissions on LP-ABS with reduced transmit power. The percentages of normal DL subframes, normal UL subframes and LP-ABS in each transmission cycle are $\alpha^{n,d},\alpha^{n,u}$ and $\alpha^{l}$, respectively.

In order to alleviate severe MBS-to-SBS interference on normal subframes, SBSs keep synchronous transmissions with the MBS on normal subframes. Besides, for the purpose of accommodating DL/UL traffic asymmetry in small cells, each SBS dynamically configures the percentages of DL and UL subframes on LP-ABS. The configuration of $(\alpha^{n,d},\alpha^{n,u},\alpha^{l})$ will be detailed described in Sec.III-A.

\subsection{Interference Model}
Let $P_{j}^{n}$ and $P_{j}^{l}$ denote the transmit power of BS $j$ on normal subframes and LP-ABS, respectively. For SBS $j\in\mathcal{S}\backslash\{0\}$, $P_{j}^{n}=P_{j}^{l}$. For the MBS, $P_{0}^{l}\le P_{0}^{n}$. In this paper, we assume that $P_{0}^{l}$ can take a value from a finite set of power levels, e.g., $P_{0}^{l}\in\{P_0,P_0+\triangle P,P_0+2\triangle P,\ldots,P_{0}^{n}\}$. Similar assumption has been adopted in \cite{M.Bennis}.
During the uplink open loop power control \cite{S.Singh}, the Signal to Noise Ratio (SNR) at the receiving BS is a fixed value. Let $\gamma$ denote the SNR at BS. The desired transmit power of UE $i$ can be expressed as $P_i^u=min\{\gamma\sigma^2/h_{i,S_i},P_{i,max}^u\}$ \cite{B.Yu}, where $P_{i,max}^u$ is the maximal transmit power of UE $i$, $\sigma^2$ is the constant additive noise power, and $h_{i,S_i}$ denote the channel gain between UE $i$ and the associated BS.

The Signal to Interference and Noise Ratio (SINR) for DL transmission to UE $i$ is given by
\begin{equation*}
  \gamma_{i}^{x,d}=\frac{P_{S_i}^x h_{i,S_i}}{\sum_{k\in \mathcal{S}\backslash\{S_i\}}P_{k}^x h_{i,k}+\sigma^2},
\end{equation*}
where $x$= `n' and `l' stand for normal subframes and LP-ABS respectively, and $h_{i,k}$ denotes the channel gain between UE $i$ and the BS $k$. Here long term achievable rate of each UE is considered.

When a UE is scheduled on normal UL subframes, interference sources may change with scheduling cycles due to scheduling dynamics \cite{D.Liu_VTC}. The precise interference level to a UE is difficult to identify. Thus, we focus on average interference to UEs. The average UL interference from neighbour cell $k$ to UEs in cell $j$ is $I_{k,j}=(\sum_{i\in\mathcal{M}_{k}}P_i^u h_{i,j})/|\mathcal{M}_k|$, where $|\mathcal{M}_k|$ is the number of UEs associated with BS $k$. The link gain is symmetric due to channel reciprocity in TDD systems \cite{P.Tarasak}, then we have $h_{i,j}=h_{j,i}$. The UL SINR of UE $i$ when it's scheduled on normal UL subframes is
\begin{equation*}
  \gamma_i^{n,u}=\frac{P_i^u h_{i,S_i}}{\sum_{k\in \mathcal{S}\backslash\{S_i\}}I_{k,S_i}+\sigma^2}.
\end{equation*}
The UL SINR of UE $i$ when it's scheduled on LP-ABS can be approximately expressed as
\begin{equation*}
\gamma_i^{l,u}=
\left\{
\begin{array}{c}
\frac{P_i^u h_{i,S_i}}{\sum_{k\in \mathcal{S}\backslash\{S_i\}}P_k^l h_{k,S_i}+\sigma^2},S_i\in\mathcal{S}\backslash\{0\};\\
0,S_i\in\{0\}.
\end{array}
\right.
\end{equation*}

Based on the above SINRs, the spectral efficiency for UE $i$ can be formulated as
\begin{equation*}
  c_i^{x,y}=log_2(1+\gamma_{i}^{x,y}),
\end{equation*}
where $x$= `n' and `l' stand for normal subframes and LP-ABS respectively, $y$= `d' and `u' stand for DL and UL respectively.

For each user, the service rate is directly determined by the spectral efficiency. Let $\textbf{s}_i=(s_i^{n,d},s_i^{l,d},s_i^{n,u},s_i^{l,u})$ denote the percentages of normal/LP-ABS subframes in a transmission cycle allocated to UE $i$ for DL/UL transmissions. The service rate of UE $i$ for DL or UL transmissions can be formulated as
\begin{equation*}
  R_i^y=B\cdot(s_i^{n,y} c_i^{n,y}+s_i^{l,y} c_i^{l,y}),y=\text{`d' or `u'},
\end{equation*}
where $B$ is the system bandwidth\footnote{For expression simplicity, $B$ is omitted in the following equations. Therefore, $R_i^y=s_i^{n,y} c_i^{n,y}+s_i^{l,y} c_i^{l,y}$ can be viewed as the normalized achievable rate of UE $i$}.

\subsection{Utility Functions for Mixed Traffic Flows}
Generally, traffic flows can be divided in two types, i.e., BE traffic flows and soft QoS traffic flows \cite{L.Chen}. Utility functions have been widely applied to model the satisfaction levels of traffic flows \cite{L.Chen}. The BE traffic flows are rather tolerant of delay and can adapt to the allocated resource. The utility function of a BE traffic flow is usually defined as a concave function of its service rate. However, the soft QoS traffic flows have intrinsic service rate requirements. When the service rate obtained by a traffic flow is less than the critical value for its service rate requirement, it will have a high-priority for resource. Otherwise, its priority for resource is low. Usually, the utility function of a soft QoS traffic flow is defined as a sigmoid function of its service rate.

We define the utility functions for a soft QoS traffic flow and a BE traffic flow as follows:
\begin{equation}\label{utility_of_qos}
  U^Q(R)=\left\{
  \begin{array}{c}
    (1-p_1)e^{q_1(R-R_{th})},R<R_{th};\\
    1-p_1e^{-q_1(R-R_{th})},R\ge R_{th};
  \end{array}
  \right.
\end{equation}
\begin{equation}\label{utility_of_be}
  U^B(R)=p_2(1-e^{-q_2R}),R\ge 0,
\end{equation}
where $p_1,q_1,p_2,q_2$ affect the slopes of the curves, $R$ is the service rate, and $R_{th}$ is the service rate requirement of soft QoS traffic.\footnote{In fact, the rate requirements for DL and UL traffic flows can be different. For the sake of simplicity, the same rate requirement is adopted. Besides, other types of utility functions can also be applied.} Similar utility functions have been adopted in \cite{L.Tan}. The derivative of a utility function $u(R)=\frac{dU(R)}{dR}$ is called as marginal utility function.

We assume that each UE can generate at most one DL and one UL traffic flow, and traffic flows are generated independently in DL and UL directions. Let $\mathcal{M}_d$ and $\mathcal{M}_u$ denote the set of DL traffic flows and UL traffic flows, respectively. Furthermore, we denote the DL traffic flows and UL traffic flows accommodated by BS $j$ by $\mathcal{M}_{j,d}$ and $\mathcal{M}_{j,u}$, respectively.

\section{PROBLEM FORMULATION AND ANALYSIS}
The introduction of LP-ABS makes the resource allocation problem in the dynamic TDD HetNet under mixed traffic complicated.
To tackle such problem, we first perform BS-level resource allocation to determine the percentages of different types of subframes. Then, based on the results of the BS-level resource allocation, we formulate the resource allocation problem at UE-level as a NUM problem.
The problem is non-convex and is complicated to solve directly. However, under given LP-ABS power, the problem can be divided into $N_s+1$ independent cell utility maximization problems. Fortunately, each cell utility maximization problem can be solved through a method of two stage decomposition, which is based on the necessary condition of its optimal solution. Finally, a resource allocation algorithm with polynomial complexity is designed.

\subsection{BS-level Resource Allocation}
The percentages of different types of subframes will affect the resource budgets of each BS for its DL and UL transmissions. Thus, such percentages should be determined before we consider resource allocation under mixed traffic. Due to numerous associated UEs per macro cell, the aggregated macro cell traffic dynamics can be averaged out \cite{M.Ding_3}. Thus, MBSs usually adopt a uniform and quasi-static configuration of DL/UL subframes and ABS periods. Besides, in our LP-ABS mechanism, each SBS keeps synchronous with the MBS on normal subframes, and applies dynamic TDD on LP-ABS. Such configuration is illustrated in Fig. \ref{frame_struc}.

We propose a BS-level resource allocation method to determine the percentages of normal DL, normal UL and LP-ABS. The purpose of the method is to obtain the maximal desired system capacity. In order to reduce complexity, we adopt a fixed LP-ABS value in this procedure. Without loss of generality, we consider that $P_0^l=0$. It should be emphasized that, by doing so, our proposed resource allocation scheme can be easily adapted to other interference mitigation mechanisms, such as the ABS based mechanism \cite{H.Ji}. In the Third Generation Partnership Project (3GPP) specification \cite{3GPP:36211}, each transmission cycle consists of $10$ subframes. We assume that at most $T\le10$ subframes can be configured as LP-ABS. Hence, $\alpha_l\in\{0,\frac{1}{T},\frac{2}{T},...,1\}$. The required resource in each link direction is related with the traffic load, namely the number of traffic flows. Besides, to achieve a required rate requirement, the required resource is inverse to the spectral efficiency. Based on the above considerations, we define the ratio of normal DL subframes to normal UL subframes as
\begin{equation*}
    \frac{\alpha^{n,d}}{\alpha^{n,u}} = \frac{\sum_{i\in\mathcal{M}_{0,d}}\frac{1}{c_i^{n,d}}}{\sum_{i\in\mathcal{M}_{0,u}}\frac{1}{c_i^{n,u}}}.
\end{equation*}
Since $\alpha^{n,d}+\alpha^{n,u}+\alpha^l=1$, $\alpha^{n,d}$ and $\alpha^{n,u}$ can be further expressed as
\begin{equation}\label{nd}
     \alpha^{n,d} = \frac{\sum_{i\in\mathcal{M}_{0,d}}\frac{1}{c_i^{n,d}}}{\sum_{i\in\mathcal{M}_{0,d}}\frac{1}{c_i^{n,d}}+
     \sum_{i\in\mathcal{M}_{0,u}}\frac{1}{c_i^{n,u}}}(1-\alpha^l),
\end{equation}
\begin{equation}\label{nu}
     \alpha^{n,u} = \frac{\sum_{i\in\mathcal{M}_{0,u}}\frac{1}{c_i^{n,u}}}{\sum_{i\in\mathcal{M}_{0,d}}\frac{1}{c_i^{n,d}}+
     \sum_{i\in\mathcal{M}_{0,u}}\frac{1}{c_i^{n,u}}}(1-\alpha^l).
\end{equation}

The average capacity of the MBS can be evaluated as
\begin{equation*}
     C_0(\alpha_l) = \frac{\sum_{i\in\mathcal{M}_{0,d}}c_i^{n,d}}{|\mathcal{M}_{0,d}|}\alpha^{n,d}+
     \frac{\sum_{i\in\mathcal{M}_{0,u}}c_i^{n,u}}{|\mathcal{M}_{0,u}|}\alpha^{n,u},
\end{equation*}
and the average capacity of SBS $j$ can be evaluated as
\begin{equation*}
    \begin{aligned}
     C_j(\alpha_l) = &\frac{\sum_{i\in\mathcal{M}_{j,d}}c_i^{n,d}}{|\mathcal{M}_{j,d}|}\alpha^{n,d}+
     \frac{\sum_{i\in\mathcal{M}_{j,u}}c_i^{n,u}}{|\mathcal{M}_{j,u}|}\alpha^{n,u}\\
     &+\frac{\sum_{i\in\mathcal{M}_{j,d}}c_i^{n,d}+\sum_{i\in\mathcal{M}_{j,u}}c_i^{n,u}}{|\mathcal{M}_{j,d}|+|\mathcal{M}_{j,u}|}\alpha_l.
    \end{aligned}
\end{equation*}
In the above evaluation, time-slots are equally allocated to traffic flows. Therefore, $\alpha^l$ can be set as
\begin{equation}\label{abs_l}
    \alpha^l = \mathop{\argmax}_{t\in\{0,\frac{1}{T},\frac{2}{T},...,1\}}\left\{\sum_{j\in\mathcal{S}}C_j(t)\right\},
\end{equation}
which can result in the highest average system capacity. Besides, $\alpha^{n,d}$ and $\alpha^{n,u}$ can be set according to Eqs. (\ref{nd}) and (\ref{nu}). Based on the above results, we can formulate a NUM problem in the following subsection.

\subsection{Network Utility Maximization Problem Formulation}
We formulate a NUM problem to perform UE-level resource allocation, through configuring $P_0^l$ and $\textbf{s}_i,i\in\mathcal{M}$. The NUM problem can be expressed as follows:
\begin{equation*}
\begin{aligned}
\mathcal{P}1:\underset{\textbf{s},P_0^l}{\text{maximize}}
&\sum_{i\in\mathcal{M}_d}U_i^d(R_i^d)+\sum_{k\in\mathcal{M}_u}U_k^u(R_k^u)\\
\text{subject to}\quad
&R_i^d=s_i^{n,d}c_i^{n,d}+s_i^{l,d}c_i^{l,d}(P_0^l), \quad \forall i\in M,\\
&R_k^u=s_k^{n,u}c_k^{n,u}+s_k^{l,u}c_k^{l,u}(P_0^l), \quad \forall i\in M,\\
&0\le s_{i}^{n,d},s_{i}^{n,l},s_{i}^{u,d},s_{i}^{u,l}, \quad \forall i\in M,\\
&\text{\em{Resource budget constraints of each BS}},\\
\end{aligned}
\end{equation*}
where the goal is to maximize the total utilities for all traffic flows in the TDD HetNet, $U_i^d(R_i^d)$ is the utility of $i$th DL traffic flow with service rate $R_i^d$, and $U_k^u(R_k^u)$ is the utility of $k$th UL traffic flow with service rate $R_k^u$. The utility function $U_i^d/U_k^u$ is decided by the traffic type of flow $i/k$. The resource budget is based on the results of the BS-level resource allocation. 

Problem $\mathcal{P}1$ is a non-convex problem, and is complicated to solve directly. Since $P_0^l$ is a one-dimension variable, we can first fix $P_0^l$. Then $\mathcal{P}1$ can be divided into $N_s+1$ independent subproblems, each of which is a cell utility maximization problem aiming to maximize the sum of utilities for UEs in a cell. The utility for a UE is the sum of utilities for its traffic flows. In this paper, we consider the cell covered by BS $j$, and formulate the cell utility maximization problem as follows:
\begin{equation*}
\begin{aligned}
\mathcal{P}2:\underset{\textbf{s}}{\text{maximize}}
&\sum_{i\in \mathcal{M}_{j,d}}\{U_i^d(s_{i}^{n,d}c_{i}^{n,d}+s_{i}^{l,d}c_{i}^{l,d})\}\\
&+\sum_{k\in \mathcal{M}_{j,u}}\{U_k^u(s_{k}^{n,u}c_{k}^{n,u}+s_{k}^{l,u}c_{k}^{l,u})\}\\
\text{subject to}\quad
&C1:\sum_{i\in \mathcal{M}_{j,d}}s_{i}^{n,d}\le \alpha^{n,d}, \\
&C2:\sum_{k\in \mathcal{M}_{j,u}}s_{k}^{n,u}\le \alpha^{n,u},  \\                                                                                                                    &C3:\sum_{i\in \mathcal{M}_{j,d}}s_{i}^{l,d}+\sum_{k\in \mathcal{M}_{j,u}}s_{k}^{l,u}\le \alpha^{l}, \\
&C4:0\le s_{i}^{x,y}, \quad \forall i\in M_j.\\
\end{aligned}
\end{equation*}
In problem $\mathcal{P}2$, the optimization variable is $\textbf{s}=\{\textbf{s}_i\}_{i\in\mathcal{M}_j}$. Constraints $C1$, $C2$ and $C3$ are resource budget constraints of BS $j$, and constraint $C4$ means that the amount of resource allocated to traffic flow $i$ should not be negative. In constraint $C4$ and the following content, $x\in\{$`n',`l'$\}$ and $y\in\{$`d',`u'$\}$. 
When UE $i\in\mathcal{M}_j$ has no DL (UL) traffic flow, $s_i^{n,d}=s_i^{l,d}=0$ ($s_i^{n,u}=s_i^{l,u}=0$).

\subsection{Solution of Cell Utility Maximization Problem $\mathcal{P}2$}
The utility function of a DL/UL traffic flow will take Eq. (\ref{utility_of_qos}) or (\ref{utility_of_be}) according to its traffic type. The objective function of $\mathcal{P}2$ is the summation of some utility functions, each of which is a concave function or a sigmoid function. Thus, the objective function is neither convex nor concave. In the following, we attempt to solve $\mathcal{P}2$ directly with low complexity.

We can see that an optimization problem has the following proposition, no matter it's a convex problem or not.
\begin{proposition} In all the solutions which satisfy the Karush-Kuhn-Tucker (KKT) \cite{convex_optimization} conditions of an optimization problem, the one optimizing the objective is the optimal solution of the original problem.
\end{proposition}

\emph{Proof:} For an optimization problem, the KKT conditions are necessary conditions of the optimal solution \cite{convex_optimization}. In other words, the optimal solution must satisfy the KKT conditions. Therefore, we can shrink the feasible solution set of problem $\mathcal{P}2$ to the set of solutions satisfying the KKT conditions. Thus we complete the proof.$\qed$

Based on \emph{Proposition 1}, we focus our attention on the solutions which satisfy the KKT conditions of problem $\mathcal{P}2$. The Lagrange function of problem $\mathcal{P}2$ is constructed as follows:
\begin{equation}\label{Lagrange}
\begin{aligned}
&\mathcal{L}(\textbf{s},\lambda,\beta,\gamma)=
-\sum_{i\in \mathcal{M}_{j,d}}\{U_i^d(R_i^d)\}-\sum_{k\in \mathcal{M}_{j,u}}\{U_k^u(R_k^u)\}\\
&+\lambda(\sum_{i\in \mathcal{M}_{j,d}}s_i^{n,d}-\alpha^{n,d})+\beta(\sum_{k\in \mathcal{M}_{j,u}}s_k^{n,d}-\alpha^{n,u})\\
&+\gamma(\sum_{i\in \mathcal{M}_{j,d}}s_i^{l,d}+\sum_{k\in \mathcal{M}_{j,u}}s_k^{l,u}-\alpha^{l})+\sum_{i\in \mathcal{M}_{j}}s_i^{x,y}v_i^{x,y},
\end{aligned}
\end{equation}
where $\lambda\ge0,\beta\ge 0,\gamma\ge 0$ and $v_i^{x,y}\ge 0$ are the Lagrange multipliers associated with constraints $C1,C2,C3$ and $C4$, respectively.
The KKT conditions can be expressed as
\begin{eqnarray}
\frac{\partial \mathcal{L}}{\partial s_i^{x,y}}=0, \quad \forall i\in M_j,\label{Lagrange1}\\
s_i^{x,y}v_i^{x,y} = 0, \quad \forall i\in M_j,\label{Lagrange2}\\
\lambda(\sum_{i\in \mathcal{M}_{j,d}}s_i^{n,d}-\alpha^{n,d})=0,\\
\beta(\sum_{k\in \mathcal{M}_{j,u}}s_k^{n,u}-\alpha^{n,u})=0,\\
\gamma(\sum_{i\in \mathcal{M}_{j,d}}s_i^{l,d}+\sum_{k\in \mathcal{M}_{j,u}}s_k^{l,u}-\alpha^{l})=0.
\end{eqnarray}
According to the above KKT conditions, we have an interesting property of the optimal solution given in \emph{Lemma 1}, which is related with UEs' DL transmissions.

\begin{lemma}
For problem $\mathcal{P}2$ and UE $i$ which obtains DL service from BS $j$, there exists Lagrange multipliers $\lambda$ and $\gamma$. If $\frac{c_i^{n,d}}{c_i^{l,d}}>\frac{\lambda}{\gamma}$, then $s_i^{n,d}>0,s_i^{l,d}=0$; else if $\frac{c_i^{n,d}}{c_i^{l,d}}<\frac{\lambda}{\gamma}$, then $s_i^{n,d}=0,s_i^{l,d}>0$.
\end{lemma}
\emph{Proof:} Substituting Eq. (\ref{Lagrange}) into Eq. (\ref{Lagrange1}), we have the following equations:
\begin{eqnarray*}
-\frac{\partial U_i^d}{\partial R_i^d}\cdot c_i^{n,d} + \lambda + v_i^{n,d} = 0,\\
-\frac{\partial U_i^d}{\partial R_i^d}\cdot c_i^{l,d} + \gamma + v_i^{l,d} = 0.
\end{eqnarray*}
Furthermore, we have
\begin{equation*}
\frac{c_i^{n,d}}{c_i^{l,d}}=\frac{\lambda+v_i^{n,d}}{\gamma+v_i^{l,d}}.
\end{equation*}
According to Eq. (\ref{Lagrange2}), it can be found that $v_i^{n,d}=0$ is equivalent to $s_i^{n,d} \ge 0$, and $v_i^{n,d}>0$ is equivalent to $s_i^{n,d}=0$. Similarly, $v_i^{l,d}=0$ is equivalent to $s_i^{l,d} \ge 0$, and $v_i^{l,d}>0$ is equivalent to $s_i^{l,d}=0$.

Since UE $i$ obtains DL service from BS $j$, we have $R_i^d=c_i^{n,d}s_i^{n,d}+c_i^{l,d}s_i^{l,d}>0$, which means that $s_i^{n,d}+s_i^{l,d} > 0$ and $s_i^{n,d},s_i^{l,d} \ge 0$. In other words, $v_i^{n,d}$ and $v_i^{l,d}$ cannot be both larger than $0$.
If $\frac{c_i^{n,d}}{c_i^{l,d}}>\frac{\lambda}{\gamma}$, we have $v_i^{n,d}>0$ and $v_i^{l,d}=0$, and we can further deduce that $s_i^{n,d}=0$ and $s_i^{l,d}>0$.
Else if $\frac{c_i^{n,d}}{c_i^{l,d}}<\frac{\lambda}{\gamma}$, we have $v_i^{n,d}=0$ and $v_i^{l,d}>0$, and further deduce that $s_i^{n,d}>0$ and $s_i^{l,d}=0$. Thus we complete the proof.
$\qed$

In addition, we have \emph{Lemma 2}, which is related with UEs' UL transmissions.
\begin{lemma}
For problem $\mathcal{P}2$ and UE $i$ which obtains UL service from BS $j$, there exists Lagrange multipliers $\beta$ and $\gamma$. If $\frac{c_i^{n,u}}{c_i^{l,u}}>\frac{\beta}{\gamma}$, then $s_i^{n,u}>0,s_i^{l,u}=0$; else if $\frac{c_i^{n,u}}{c_i^{l,u}}<\frac{\beta}{\gamma}$, then $s_i^{n,u}=0,s_i^{l,u}>0$.
\end{lemma}
\emph{Proof:} The proof is similar to that of \emph{Lemma 1}.
$\qed$

\emph{Lemma 1} can be further explained as follows. BS $j$ can be viewed as two virtual BSs, $j1$ and $j2$. Virtual BS $j1$ has only normal DL subframes, and virtual BS $j2$ has LP-ABS. Parameters $c_i^{n,d}$ and $\frac{\lambda}{\gamma}c_i^{l,d}$ can be seen as the spectral efficiencies when UE $i$ is associated with virtual BS $j1$ and $j2$, respectively. Thus, each UE will associate with the virtual BS, from which the UE can obtain higher spectral efficiency. We can find that the resource allocation problem $\mathcal{P}2$ is equivalent to the user association problem in \cite{P.Xue} or \cite{Q.Ye}. According to \emph{Proposition $3$} in \cite{Q.Ye}, there is at most one UE associated with both $j1$ and $j2$. In this paper, we assume there are no UEs in BS $j$ associate to both virtual BSs. That is to say, no UEs satisfy $s_i^{n,d}>0,s_i^{l,d}>0$. However, based on the results of \cite{P.Xue}, such assumption almost does not affect the performance of UEs. Thus such assumption is feasible.

\begin{figure}[t]
  \centering
  \includegraphics[width=0.4\textwidth]{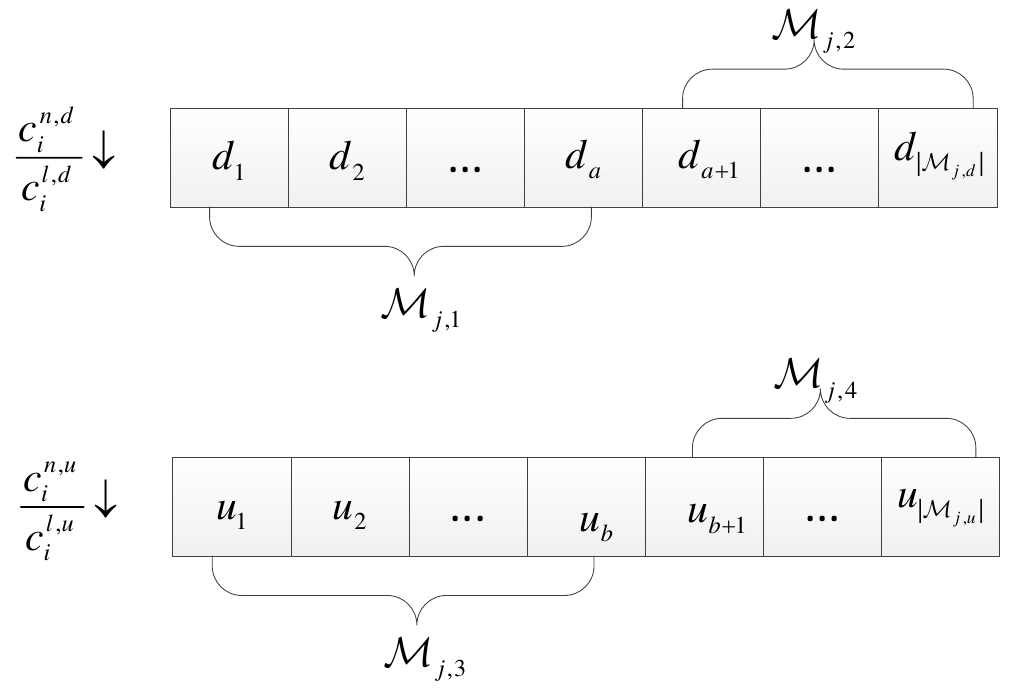}\\
  \caption{Sorted UEs of $\mathcal{M}_j$ according to descending order of $\frac{c_i^{n,d}}{c_i^{l,d}}$ and $\frac{c_i^{n,u}}{c_i^{l,u}}$}\label{sort_ue}
\end{figure}

Let BS $j$ sort the DL traffic flows of UEs in $\mathcal{M}_j$, i.e., $\mathcal{M}_{j,d}$, according to the descending order of $\frac{c_i^{n,d}}{c_i^{l,d}}$. The sorted traffic flows are illustrated in Fig. \ref{sort_ue}, and $d_a$ is the index of $a$th DL traffic flow. Besides, let BS $j$ sort the UL traffic flows of UEs in $M_j$, i.e., $\mathcal{M}_{j,u}$, according to the descending order of $\frac{c_i^{n,u}}{c_i^{l,u}}$, and $u_b$ is the index of $b$th UL traffic flow. Given $a\in\{1,2,\ldots,|\mathcal{M}_{j,d}|-1\}$ and $b\in\{1,2,\ldots,|\mathcal{M}_{j,u}|-1\}$, we can construct four sets of traffic flows, i.e., $\mathcal{M}_{j,1}$, $\mathcal{M}_{j,2}$, $\mathcal{M}_{j,3}$ and $\mathcal{M}_{j,4}$, as shown in Fig. \ref{sort_ue}.

Based on \emph{Proposition 1}, \emph{Lemma 1} and \emph{2}, under the optimal solution of $\mathcal{P}2$, there exist specific $a$ and $b$ such that traffic flows in $\mathcal{M}_{j,1}$, $\mathcal{M}_{j,2}$, $\mathcal{M}_{j,3}$ and $\mathcal{M}_{j,4}$ only obtain resource from normal DL subframes, LP-ABS, normal UL subframes and LP-ABS of BS $j$, respectively. In this work, we propose a method of two stage decomposition for solving $\mathcal{P}2$. In the first stage of decomposition, we fix $a,b$, and divide $\mathcal{P}2$ into three independent subproblems. Then in the second stage of decomposition, under certain conditions, the optimal solution of each subproblem can be divided into two cases. Finally, we can traverse and find the specific $a,b$ that maximize the objective function of $\mathcal{P}2$, as well as the optimal solution of $\mathcal{P}2$. The detailed procedures will be analyzed in the following content.

Based on the above analysis, we can perform the first stage of decomposition. Under fixed $a$ and $b$, we can construct the following three independent subproblems.
\begin{equation*}
\begin{aligned}
\mathcal{P}2.1:\underset{\textbf{s}^{n,d}}{\text{maximize}}
&\sum_{i\in \mathcal{M}_{j,1}}U_i^d(s_{i}^{n,d}c_{i}^{n,d})\\
\text{subject to}\quad
&\sum_{i\in \mathcal{M}_{j,1}}s_{i}^{n,d}\le \alpha^{n,d}, \\
&0\le s_{i}^{n,d}, \quad \forall i\in \mathcal{M}_{j,1}.\\
\end{aligned}
\end{equation*}
\begin{equation*}
\begin{aligned}
\mathcal{P}2.2:\underset{\textbf{s}^{n,u}}{\text{maximize}}
&\sum_{i\in \mathcal{M}_{j,3}}U_i^u(s_{i}^{n,u}c_{i}^{n,u})\\
\text{subject to}\quad
&\sum_{i\in \mathcal{M}_{j,3}}s_{i}^{n,u}\le \alpha^{n,u}, \\
&0\le s_{i}^{n,u}, \quad \forall i\in \mathcal{M}_{j,3}.\\
\end{aligned}
\end{equation*}
\begin{equation*}
\begin{aligned}
\mathcal{P}2.3:\underset{\textbf{s}^{l,d},\textbf{s}^{l,u}}{\text{maximize}}
&\sum_{i\in \mathcal{M}_{j,2}}U_i^d(s_{i}^{l,d}c_{i}^{l,d})+\sum_{k\in \mathcal{M}_{j,4}}U_i^u(s_{k}^{l,u}c_{k}^{l,u})\\
\text{subject to}\quad
&\sum_{i\in \mathcal{M}_{j,2}}s_{i}^{l,d}+\sum_{k\in \mathcal{M}_{j,4}}s_{k}^{l,u}\le \alpha^{l}, \\
&0\le s_{i}^{l,d},s_{k}^{l,u}, \quad \forall i\in \mathcal{M}_{j,2},\forall k\in \mathcal{M}_{j,4}.\\
\end{aligned}
\end{equation*}

It should be emphasized that the first stage of decomposition is independent of traffic types. The above three subproblems are corresponding to the problems of allocating normal DL subframes of BS $j$ to $\mathcal{M}_{j,1}$, allocating normal UL subframes of BS $j$ to $\mathcal{M}_{j,3}$, and allocating LP-ABS of BS $j$ to $\mathcal{M}_{j,2}\cup\mathcal{M}_{j,4}$, respectively. Since the above three subproblems have identical structures, we focus on subproblem $\mathcal{P}2.1$.

\begin{figure}
  \centering
  \includegraphics[width=0.4\textwidth]{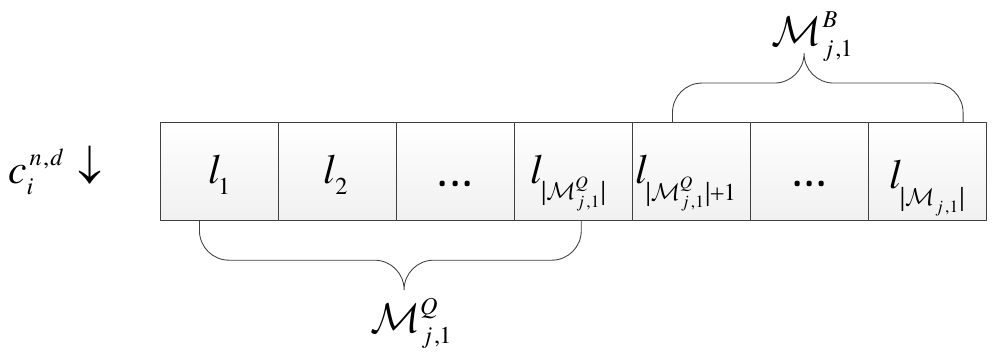}\\
  \caption{Queue of traffic flows in $\mathcal{M}_{j,1}$, which is sorted according to their traffic types and the descending order of $c_i^{n,d}$ }\label{ue_queue}
\end{figure}

As for the traffic flows in $\mathcal{M}_{j,1}$, some are soft QoS traffic flows, and the others are BE traffic flows. The traffic flows in $\mathcal{M}_{j,1}$ could be divided into two parts, as illustrated in Fig. \ref{ue_queue}. The first part includes the soft QoS traffic flows and is followed by the second part, which includes the BE traffic flows. In each part, the traffic flows are sorted according to the descending order of $c_i^{n,d}$. The number in each box means the index of a traffic flow. Under the premise of not causing confusion, we will use indexes $1,2,...,|\mathcal{M}_{j,1}|$ instead of $l_1,l_2,...,l_{|\mathcal{M}_{j,1}|}$.

Let $\mathcal{M}_{j,1}^Q$ and $\mathcal{M}_{j,1}^B$ denote the sets of the sorted traffic flows in the two parts. Subproblem $\mathcal{P}2.1$ is equivalent to the following problem.
\begin{equation*}
\begin{aligned}
\mathcal{P}3:\underset{\textbf{s}^{n,d}}{\text{maximize}}
&\sum_{i\in \mathcal{M}_{j,1}^Q}U_i^Q(s_{i}^{n,d}c_{i}^{n,d})+\sum_{k\in \mathcal{M}_{j,1}^B}U_i^B(s_{k}^{n,d}c_{k}^{n,d})\\
\text{subject to}\quad
&\sum_{i\in \mathcal{M}_{j,1}}s_{i}^{n,d}\le \alpha^{n,d}, \\
&0\le s_{i}^{n,d}, \quad \forall i\in \mathcal{M}_{j,1},\\
\end{aligned}
\end{equation*}
where the optimization variable is $\textbf{s}^{n,d}=\{s_i^{n,d}\}_{i\in\mathcal{M}_j}$. It should be noted that $s_{i}^{n,d}=0$ if $i\notin \mathcal{M}_{j,1}$.

The marginal utility function of a BE traffic flow is
\begin{equation}
u^B(R)=p_2q_2e^{-q_2R},R\ge 0.
\end{equation}
The function decreases as $R$ increases. Let $u_i^B(.)^{-1}$ be the inverse function of $u_i^B(.)$. In addition, the marginal utility function of a soft QoS traffic flow is
\begin{equation}
u^Q(R)=
\left\{
\begin{array}{c}
(1-p_1)q_1e^{q_1(R-R_{th})},R<R_{th};\\
p_1q_1e^{-q_1(R-R_{th})},R\ge R_{th}.
\end{array}
\right.
\end{equation}
When $R < R_{th}$, the function increases as $R$ increases. However, when $R \ge R_{th}$, the function decreases as $R$ increases. When $R_i^d < R_{th}$, let $u_i^Q(.)^{-1}$ be the inverse function of $u_i^Q(.)$. When $R_i^d \ge R_{th}$, we define $\widehat u_i(R_i)=u_i^Q(R_i+R_{th})$, and denote $\widehat u_i(.)^{-1}$ the inverse function of $\widehat u_i(.)$. For problem $\mathcal{P}3$, we have the following lemma, which presents a necessary condition for the optimal solution of $\mathcal{P}3$.

\begin{lemma}\label{SN_condition}
Let $\textbf{s}^{n,d}$ be the optimal solution of $\mathcal{P}3$, there exists a Lagrange multiplier $\lambda\ge0$ such that
\begin{equation}\label{th4_eq1}
c_i^{n,d}u_i^Q(c_i^{n,d}s_i^{n,d})=c_k^{n,d}u_k^B(c_k^{n,d}s_k^{n,d})=\lambda,\forall s_i^{n,d},s_k^{n,d}>0,
\end{equation}
\begin{equation}\label{th4_eq2}
\underset{i\in\mathcal{M}_{j,1}}{\sum}s_i^{n,d}=\alpha^{n,d},
\end{equation}
and there is at most one traffic flow satisfying $\frac{du_i(R_i)}{dR_i}>0$ in $\mathcal{M}_{j,1}$. Besides, it must be the last allocated traffic flow.
\end{lemma}
\emph{Proof:}
If $\textbf{s}^{n,d}$ is the optimal solution of $\mathcal{P}3$, then $\textbf{s}^{n,d}$ satisfies the KKT conditions of $\mathcal{P}3$, which are equivalent to Eqs. (\ref{th4_eq1}) and (\ref{th4_eq2}).

Since for an arbitrary BE traffic flow $k$, we always have $\frac{du_k(R_k)}{dR_k}<0$. Therefore, only soft QoS traffic flows may satisfy $\frac{du_i(R_i)}{dR_i}>0$. We assume that soft QoS traffic flows $i$ and $j$ satisfy $\frac{du_i(R_i)}{dR_i}>0$ and $\frac{du_j(R_j)}{dR_j}>0$, respectively. Consider another resource allocation solution $\widehat{\textbf{s}}^{n,d}$, in which the percentages of normal DL subframes allocated to flow $i$ and $j$ are $s_i^{n,d}-\triangle s$ and $s_j^{n,d}+\triangle s$, respectively. Since the utility function in Eq. (\ref{utility_of_qos}) is continuous and increasing, the difference in the objective of $P3$ when $\textbf{s}^{n,d}$ and $\widehat{\textbf{s}}^{n,d}$ are adopted can be expressed as
\begin{equation*}
\begin{aligned}
&[U_i^Q(c_i^{n,d}(s_i^{n,d}-\triangle s))+U_j^Q(c_j^{n,d}(s_j^{n,d}+\triangle s))]\\
&-[U_i^Q(c_i^{n,d}s_i^{n,d})+U_j^Q(c_j^{n,d}s_j^{n,d})]\\
&=\int_{s_j^{n,d}}^{s_j^{n,d}+\triangle s}{c_j^{n,d}u_j^Q(c_j^{n,d}s)ds}-\int_{s_i^{n,d}-\triangle s}^{s_i^{n,d}}{c_i^{n,d}u_i^Q(c_i^{n,d}s)ds}\\
&>c_j^{n,d}u_j^Q(c_j^{n,d}s_j^{n,d})-c_i^{n,d}u_i^Q(c_i^{n,d}s_i^{n,d})\\
&=0.
\end{aligned}
\end{equation*}
Such result violates the condition that $\textbf{s}^{n,d}$ is the optimal solution of $\mathcal{P}3$.

The traffic flows in $\mathcal{M}_{j,1}^Q$ and $\mathcal{M}_{j,1}^B$ have been sorted according to the descending order of $c_i^{n,d}$. In $\textbf{s}^{n,d}$, if $s_i^{n,d}=s'>0$ and $s_k^{n,d}=0, k < i \le |{M}_{j,1}^Q|$, we can always find another resource allocation solution $\widehat{\textbf{s}}^{n,d}$, in which $s_{l_k}^{n,d}=s'$ and $s_{l_i}^{n,d}=0$. It can be found that $\widehat{\textbf{s}}^{n,d}$ can further improve the objective function of $\mathcal{P}3$, which is contradict to the optimality of $\textbf{s}^{n,d}$. Therefore, if there exists a traffic flow satisfying $\frac{du_i(R_i)}{dR_i}>0$, it must be the last allocated soft QoS traffic flow in $\mathcal{M}_{j,1}^Q$. Thus we complete the proof.
$\qed$

As for a soft QoS traffic flow $i$, $\frac{du_i(R_i)}{dR_i}>0$ is equivalent to that $R_i^d < R_{th}$. \emph{Lemma \ref{SN_condition}} indicates that for the optimal solution of $\mathcal{P}3$, there is at most one soft traffic flow satisfying $R_i^d<R_{th}$, and it's the last allocated traffic flow in $\mathcal{M}_{j,1}^Q$. According to \emph{Lemma \ref{SN_condition}}, we can perform the second stage of decomposition. We first assume $m\le |\mathcal{M}_{j,1}^Q|$ soft QoS traffic flows obtain service from BS $j$, and then the optimal solution of $\mathcal{P}3$ will belong to the following two cases.
\subsubsection{Case 1 ($R_m^d\ge R_{th}$)} Based on Eq. (\ref{th4_eq2}), we have
\begin{equation}\label{case1}
\begin{aligned}
\sum_{i=1}^{m}[(\widehat u_i(\frac{\lambda}{c_i^{n,d}})^{-1}+R_{th})/c_i^{n,d}]+\sum_{k\in \mathcal{M}_{j,1}^B}u_k^B(\frac{\lambda}{c_k^{n,d}})^{-1}/c_k^{n,d}\\
=\alpha^{n,d}.
\end{aligned}
\end{equation}
Let $\lambda_1$ be the solution of Eq. (\ref{case1}). Then based on Eq. (\ref{th4_eq1}), we can obtain the optimal solution of $\mathcal{P}3$ as follows:
\begin{equation}\label{s_in_case1}
s_i^{n,d}=
\left\{
\begin{array}{c}
(\widehat u_i(\frac{\lambda_1}{c_i^{n,d}})^{-1}+R_{th})/c_i^{n,d},\quad i=1,2,\ldots,m;\\
0,\quad i=m+1,\ldots,|\mathcal{M}_{j,1}^Q|;\\
u_i^B(\frac{\lambda_1}{c_i^{n,d}})^{-1}/c_i^{n,d},\quad i=|\mathcal{M}_{j,1}^Q|+1,\ldots,|\mathcal{M}_{j,1}|.
\end{array}
\right.
\end{equation}

\subsubsection{Case 2 ($R_m^d<R_{th}$)} Based on Eq. (\ref{th4_eq2}), we have
\begin{equation}\label{case2}
\begin{aligned}
&\sum_{i=1}^{m-1}[(\widehat u_i(\frac{\lambda}{c_i^{n,d}})^{-1}+R_{th})/c_i^{n,d}]+u_m^Q(\frac{\lambda}{c_m^{n,d}})^{-1}/c_m^{n,d}\\
&+\sum_{k\in \mathcal{M}_{j,1}^B}u_k^B(\frac{\lambda}{c_k^{n,d}})^{-1}/c_k^{n,d}=\alpha^{n,d}.
\end{aligned}
\end{equation}
Let $\lambda_2$ be the solution of Eq. (\ref{case2}). Then based on Eq. (\ref{th4_eq1}), we can obtain the optimal solution of $\mathcal{P}3$ as follows:
\begin{equation}\label{s_in_case2}
s_i^{n,d}=
\left\{
\begin{array}{c}
(\widehat u_i(\frac{\lambda_2}{c_i^{n,d}})^{-1}+R_{th})/c_i^{n,d},\quad i=1,2,\ldots,m-1;\\
u_i^Q(\frac{\lambda_2}{c_i^{n,d}})^{-1}/c_i^{n,d},\quad i=m;\\
0,\quad i=m+1,\ldots,|\mathcal{M}_{j,1}^Q|;\\
u_i^B(\frac{\lambda_2}{c_i^{n,d}})^{-1}/c_i^{n,d},\quad i=|\mathcal{M}_{j,1}^Q|+1,\ldots,|\mathcal{M}_{j,1}|.
\end{array}
\right.
\end{equation}

Then, we can traverse $m$ in set $\{1,2,\ldots,|\mathcal{M}_{j,1}^Q|\}$, and find the specific $m$ and $\lambda$ that optimize the objective of $\mathcal{P}3$, as well as the optimal solution of $\mathcal{P}3$. Based on the analysis and the two stages of decomposition, we can design algorithms to solve $\mathcal{P}3$ and $\mathcal{P}2$ in the next subsection.

\subsection{Utility Based Resource Allocation Algorithm}
\begin{algorithm}[h]
\caption{Algorithm to Find Optimal Solution of \emph{P3}}
\label{Algo1}
\textbf{Initialize} $\mathcal{M}_{j,1}$, $\alpha^{n,d}$; let $U=0$;\\
sort the soft QoS traffic flows in $\mathcal{M}_{j,1}$ in descending order of $c_i^{n,d}$ and store them in $\mathcal{M}_{j,1}^Q$;\\
\For{$m=1$ to $|\mathcal{M}_{j,1}^Q|$}{
    calculate $\textbf{s}_1^{n,d}$ according to Eq. (\ref{s_in_case1});\\
    denote the objective of $P3$ in such case by $U1$;\\
    calculate $\textbf{s}_2^{n,d}$ according to Eq. (\ref{s_in_case2});\\
    denote the objective of $P3$ in such case by $U2$;\\
    \If{$U1>U$ \& $U1>U2$}{
        $U=U1,\textbf{s}^{n,d}=\textbf{s}_1^{n,d}$;
        }
    \If{$U2>U$ \& $U2>U1$}{
        $U=U2,\textbf{s}^{n,d}=\textbf{s}_2^{n,d}$;
        }
    }
return $\textbf{s}^{n,d},U$.
\end{algorithm}

Based on \emph{Lemma \ref{SN_condition}}, we design an algorithm to find the optimal solution of problem $\mathcal{P}3$ (subproblem $\mathcal{P}2.1$), as depicted in Algorithm $1$. For each $m\in\{1,2,\ldots,|\mathcal{M}_{j,1}^Q|\}$, the resource allocation result obtained from \emph{Lines $4-13$} is a solution that satisfies the necessary condition presented by \emph{Lemma 3}. Through comparing the objective of $\mathcal{P}3$ under different $m$, we can find the optimal solution of $\mathcal{P}3$.

Subproblems $\mathcal{P}2.2$ and $\mathcal{P}2.3$ have similar structures with subproblem $\mathcal{P}2.1$. Thus, we can also apply Algorithm $1$ to solve these two subproblems. However some parameters in Algorithm $1$, e.g., set of traffic flows and spectral efficiencies, should be adjusted according to the specific subproblem. For example, when solving subproblem $\mathcal{P}2.3$, we introduce a parameter $c_i^l$ $ (i\in\mathcal{M}_{j,2}\cup\mathcal{M}_{j,4})$. For a DL traffic flow $i\in\mathcal{M}_{j,2}$, let $c_i^l=c_i^{l,d}$. For an UL traffic flow $i\in\mathcal{M}_{j,4}$, let $c_i^l=c_i^{l,u}$. Then in \emph{Line $2$} of Algorithm $1$, we sort the soft traffic flows in $\mathcal{M}_{j,2}\cup\mathcal{M}_{j,4}$ according to the descending order of $c_i^l$.

In addition, we design an algorithm to solve problem $\mathcal{P}2$, which is named Utility Based Resource Allocation (UBRA) algorithm and is presented in Algorithm $2$. Under fixed $a$ and $b$, the resource allocation result obtained from \emph{Lines $7-14$} is a solution which satisfies the KKT conditions of $\mathcal{P}2$. Through comparing the objective of $\mathcal{P}2$ under different $a$ and $b$, we can find the optimal solution of $\mathcal{P}2$.
\begin{algorithm}[h]
\caption{UBRA Algorithm at BS $j$}
\label{Algo2}
\textbf{Initialize} $\mathcal{M}_{j,d}$, $\mathcal{M}_{j,u}$; let $U=0$;\\ 
\textbf{Initialize} $(\alpha_n^{n,d},\alpha_n^{n,u},\alpha^{l})$ according to the BS-level resource allocation in Sec.III-A;\\
sort the traffic flows in $\mathcal{M}_{j,d}$ in descending order of $\frac{c_i^{n,d}}{c_i^{l,d}}$;\\
sort the traffic flows in $\mathcal{M}_{j,u}$ in descending order of $\frac{c_i^{n,u}}{c_i^{l,u}}$;\\
\For{$b=1$ to $|\mathcal{M}_{j,d}|-1$}{
    \For{$a=1$ to $|\mathcal{M}_{j,u}|-1$}{
        update $\mathcal{M}_{j,1},\mathcal{M}_{j,2},\mathcal{M}_{j,3},\mathcal{M}_{j,4}$ according to the illustration of Fig. \ref{sort_ue};\\
        apply Algorithm $1$ to solve subproblem $P2.1$, and denote the results by $\textbf{s}^{n,d}$ and $U_1$;\\
        apply Algorithm $1$ to solve subproblem $P2.2$, and denote the results by $\textbf{s}^{n,u}$ and $U_2$;\\
        apply Algorithm $1$ to solve subproblem $P2.3$, and denote the results by $\textbf{s}^{l}=\{\textbf{s}^{l,d},\textbf{s}^{l,u}\}$ and $U_3$;\\
        \If{$U<U_1+U_2+U_3$}{
            $U=U_1+U_2+U_3$;\\
            calculate $\textbf{s}$ according to $\textbf{s}^{n,d},\textbf{s}^{n,u},\textbf{s}^{l,d}$ and $\textbf{s}^{l,u}$;
        }
    }
}
return $\textbf{s},U$.
\end{algorithm}

Next, we will evaluate the computational complexity of the UBRA algorithm. We focus on a special case, where $|\mathcal{M}_{j}|=|\mathcal{M}_{j,d}|=|\mathcal{M}_{j,u}|=N$. The complexities of \emph{Line $3$} and \emph{Line $4$} depend on the adopted sorting algorithm. When the basic straight insertion sort method is adopted, both the complexities of \emph{Line $3$} and \emph{Line $4$} are $O(N^2)$. Under each fixed $(a,b)$, Algorithm $1$ is applied in \emph{Line $7$}. The complexity of Algorithm $1$ is mainly decided by the sort operation in \emph{Line $2$}. We propose to utilize the sorted set $\mathcal{M}_{j,1}^Q$ under $(a-1,b)$ to reduce the complexity of sort operation under $(a,b)$. Since the set $\mathcal{M}_{j,1}^Q$ has been sorted under $(a-1,b)$, when considering the situation under $(a,b)$, we just need to insert $d_a$ into $\mathcal{M}_{j,1}^Q$ to obtain a new sorted set. Through this method, the complexity of \emph{Line $7$} can be reduced to $O(N)$. Similarly, the complexities of \emph{Line $8$} and \emph{Line $9$} are also $O(N)$. Hence, the complexity of \emph{Lines $4-15$} is $O(N^3)$. In consequence, the complexity of the UBRA algorithm is $O(N^3)$, which is polynomial.

In our proposed resource allocation scheme, the UBRA algorithm is executed under a given LP-ABS power $P_0^l$.
Each time when the MBS changes $P_0^l$, the UBRA algorithm is executed independently at each BS. The MBS can collect the utility of each SBS through X2 interface \cite{X2_interface}. After the MBS iterates $P_0^l$ in $\{P_0,P_0+\triangle P,P_0+2\triangle P,\ldots,P_{0}^{n}\}$, the optimal LP-ABS power of the MBS and the optimal resource allocation of UEs in the TDD HetNet can be obtained.

The proposed resource allocation scheme can be easily adapted to other interference mitigation mechanisms. For example, when considering the ABS based mechanism proposed in \cite{H.Ji}, we can set $P_0^l=0$ and adjust the spectral efficiencies in the algorithm accordingly. Then, Algorithm $2$ can be directly applied to perform resource allocation under such interference mitigation mechanism.

\section{NUMERICAL SIMULATIONS}
In this section, numerical results are presented to evaluate the performance of the proposed resource allocation scheme. The performance of the proposed algorithm under different interference mitigation mechanisms will be compared. Besides, the performance of the proposed resource allocation scheme under mixed traffic scenarios is evaluated.
\begin{table}[h]
\centering
\caption{Simulation parameters}
\label{parameters}
 \begin{tabular}{ccc}
 \hline
 Parameters&Macro cell&Small cell\\
 \hline
 System Bandwidth&20 MHz&20 MHz\\
 Cell Site&ISD = 500 m&Radius = 40 m\\
 Pathloss&$128.1+37.6log_{10}d$&$140.7+36.7log_{10}d$\\
 Shadowing Deviation&4 dB&4 dB\\
 Noise Power Density&-174 dBm/Hz&-174 dBm/Hz\\
 Number of BSs&1&6\\
 Transmit Power&43 dBm&30 dBm\\
 Antenna Gain&14 dBi&10 dBi\\
 Bias Value&0 dB&6 dB\\
 \hline
 \end{tabular}
\end{table}
\subsection{Simulation Setup}
In our simulation scenario, a co-channel TDD HetNet is considered, as illustrated in Fig. \ref{network_arch}. One MBS is located in the center of the macro cell, and several SBSs are randomly deployed around the MBS. In the network, $3/5$ of total UEs are located under the coverage of the SBSs, and the other UEs are randomly distributed in the macro cell. In the simulations, each UE will generate a DL traffic flow with probability $1$ and an UL traffic flow with probability of $50\%$. The target received SNR at BS from UE's UL transmissions is 10dB, and the maximal transmit power of UEs is 23dBm. Other default simulation configurations are listed in Table \ref{parameters}, which are selected based on 3GPP LTE specification \cite{3GPP:Spec}. In the configuration of pathloss, parameter $d$ means the distance from a UE to its associated BS in kilometers.

As for the utility functions, $p_1=0.2,p_2=0.4,q_1=q_2=12.8$, and the default rate requirement for soft QoS traffic flows is set as 0.5Mbps. As for the low power configuration of $P_0^l$, the minimal value $P_0=22$dBm, and $\triangle P=3$dBm.

In our simulation, at most $T=4$ subframes in a transmission cycle can be configured as LP-ABS. Let $\alpha_{n,d},\alpha_{n,u}$ and $\alpha_{l}$ denote the results of the BS-level resource allocation. Except the proposed LP-ABS mechanism, the UBRA algorithm is also evaluated under several interference mitigation mechanisms as follows.
\begin{enumerate}
  \item eICIC (ABS based) \cite{H.Ji}: the MBS configures the ratio of DL subframes, UL subframes and ABS as $\alpha_{n,d}$:$\alpha_{n,u}$:$\alpha_{l}$; each SBS configures DL, UL and dynamic TDD transmissions on the DL subframes, UL subframes and ABS of the MBS, respectively.
  \item UM-ABS \cite{Z.Shen}: the MBS configures the ratio of DL and UL subframes as $\alpha_{n,d}$:$(\alpha_{n,u}+\alpha_{l})$; each SBS configures DL and dynamic TDD transmissions on the DL and UL subframes of the MBS, respectively.
  \item Synchronous \cite{M.Ding_3}: the MBS and SBSs configure synchronous DL and UL transmissions, and the ratio of DL and UL subframes is set as $(\alpha_{n,d}+\alpha_{l})$:$\alpha_{n,u}$.
\end{enumerate}

In the following figures, `Synch' means the synchronous mechanism.
\begin{figure}[t]
\centering
\subfigure[Performance of DL throughput.]{
\begin{minipage}[b]{0.4\textwidth}
\includegraphics[width=1\textwidth]{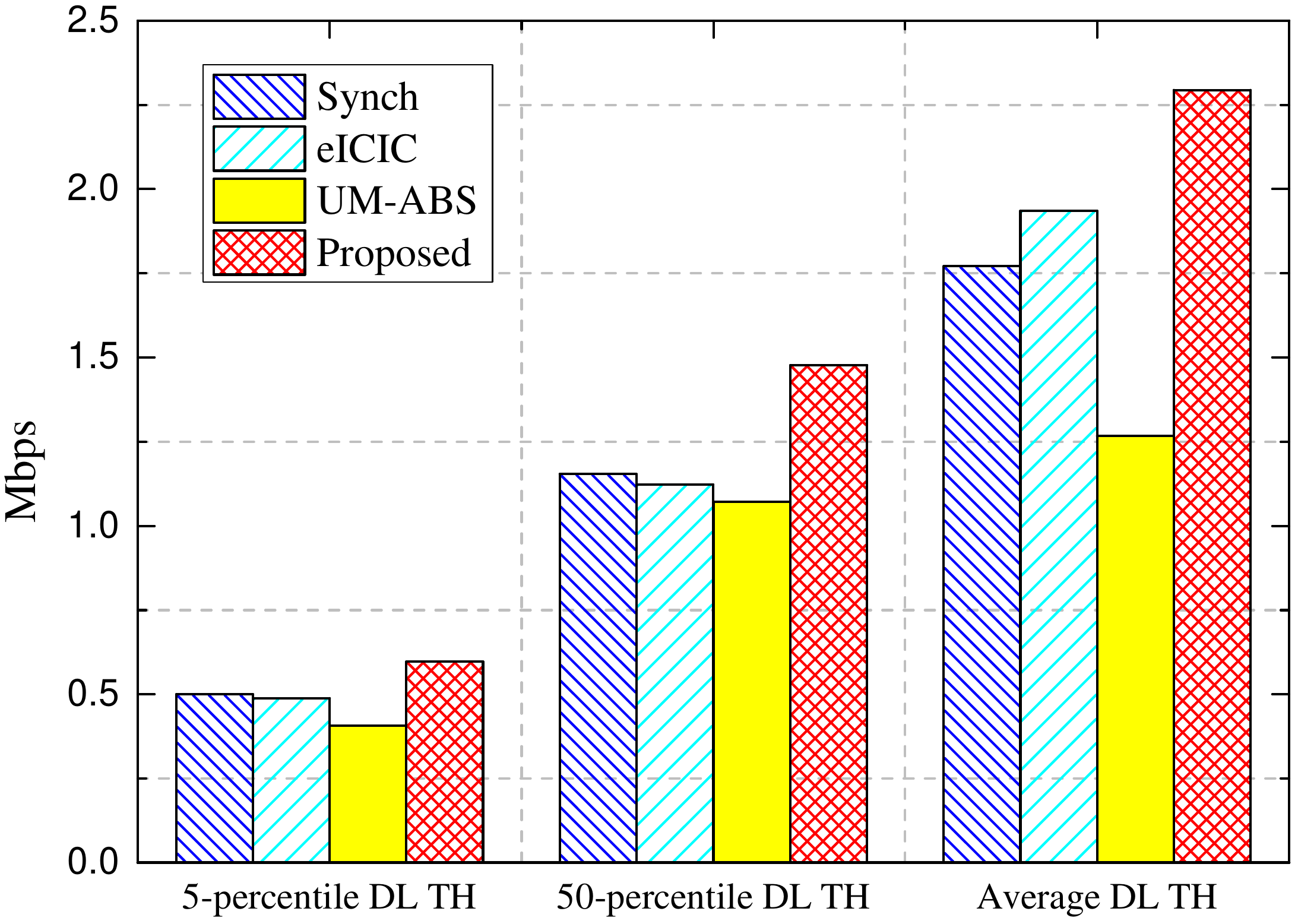}
\end{minipage}
}
\subfigure[Performance of UL throughput.]{
\begin{minipage}[b]{0.4\textwidth}
\includegraphics[width=1\textwidth]{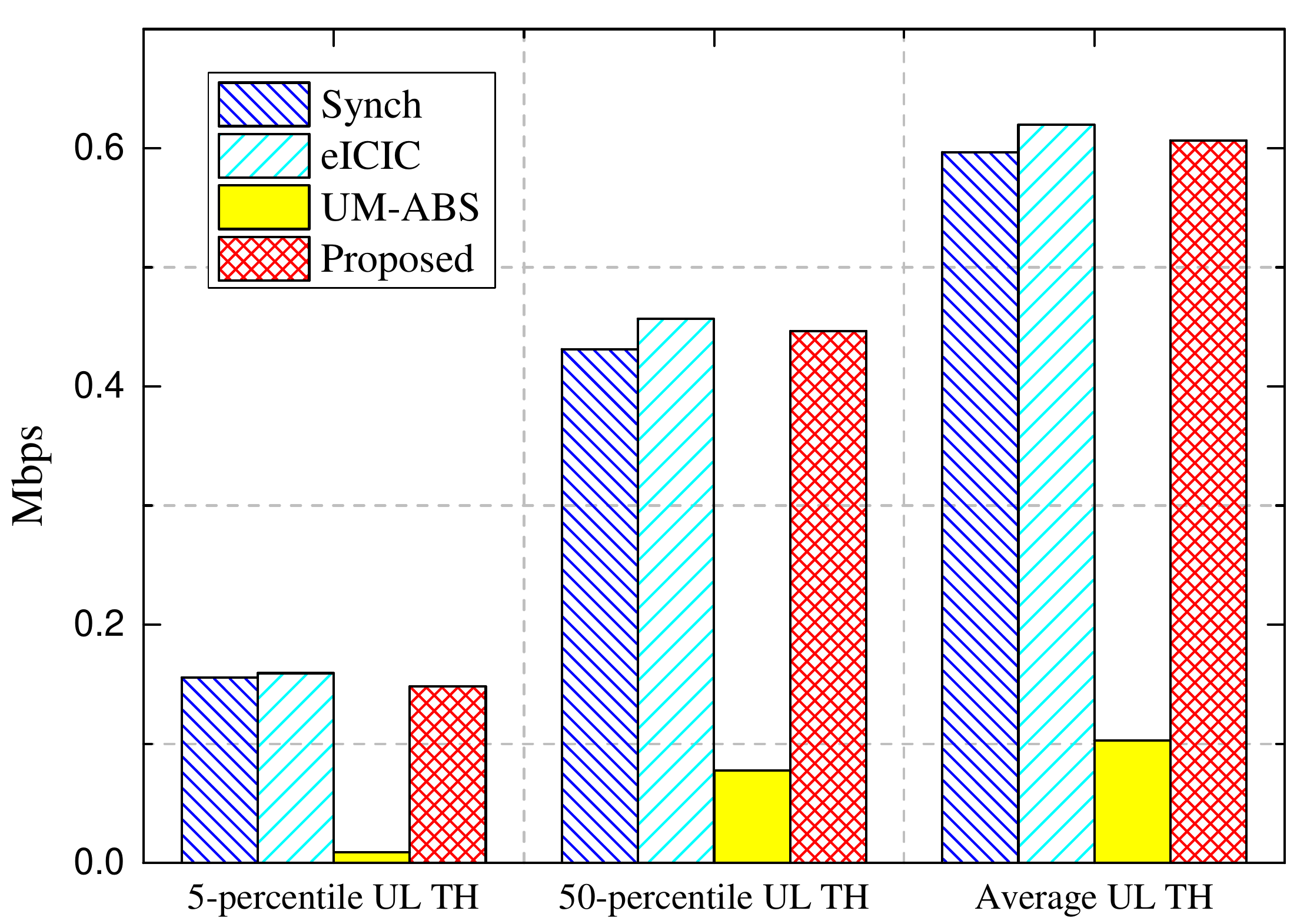}
\end{minipage}
}
\caption{Throughput performance of UEs when only BE traffic flows exist (TH = throughput).}\label{fig1}
\end{figure}
\subsection{Simulation Results}

In Fig. \ref{fig1}, we evaluate network performance in terms of the 5-percentile/50-percentile/average DL and UL throughput. In this simulation, a total of $200$ UEs are distributed in the TDD HetNet, and only BE traffic flows are generated. From Fig. \ref{fig1}(a), it can be seen that the performance of the proposed LP-ABS mechanism is superior to all the other mechanisms in the DL direction. This is due to the fact that our proposed LP-ABS mechanism can effectively utilize all transmission opportunities of the MBS. Besides, LP-ABS also takes the advantage of the eICIC mechanism, which can alleviate severe cross tier interference from macro cells to small cells. As shown in Fig. 5(b), in the UL direction, the performance of our proposed LP-ABS mechanism is quite close to that of the eICIC mechanism and the synchronous mechanism. This can be explained as follows. The above three mechanisms have almost the same amount of UL transmission resources and similar spectral efficiencies, which will result in close UL performance. Based on the above observations, the proposed LP-ABS mechanism can make a tradeoff between resource utilization and interference mitigation.

\begin{figure}[t]
\centering
\subfigure[Performance of DL throughput under different network loads.]{
\begin{minipage}[b]{0.4\textwidth}
\includegraphics[width=1\textwidth]{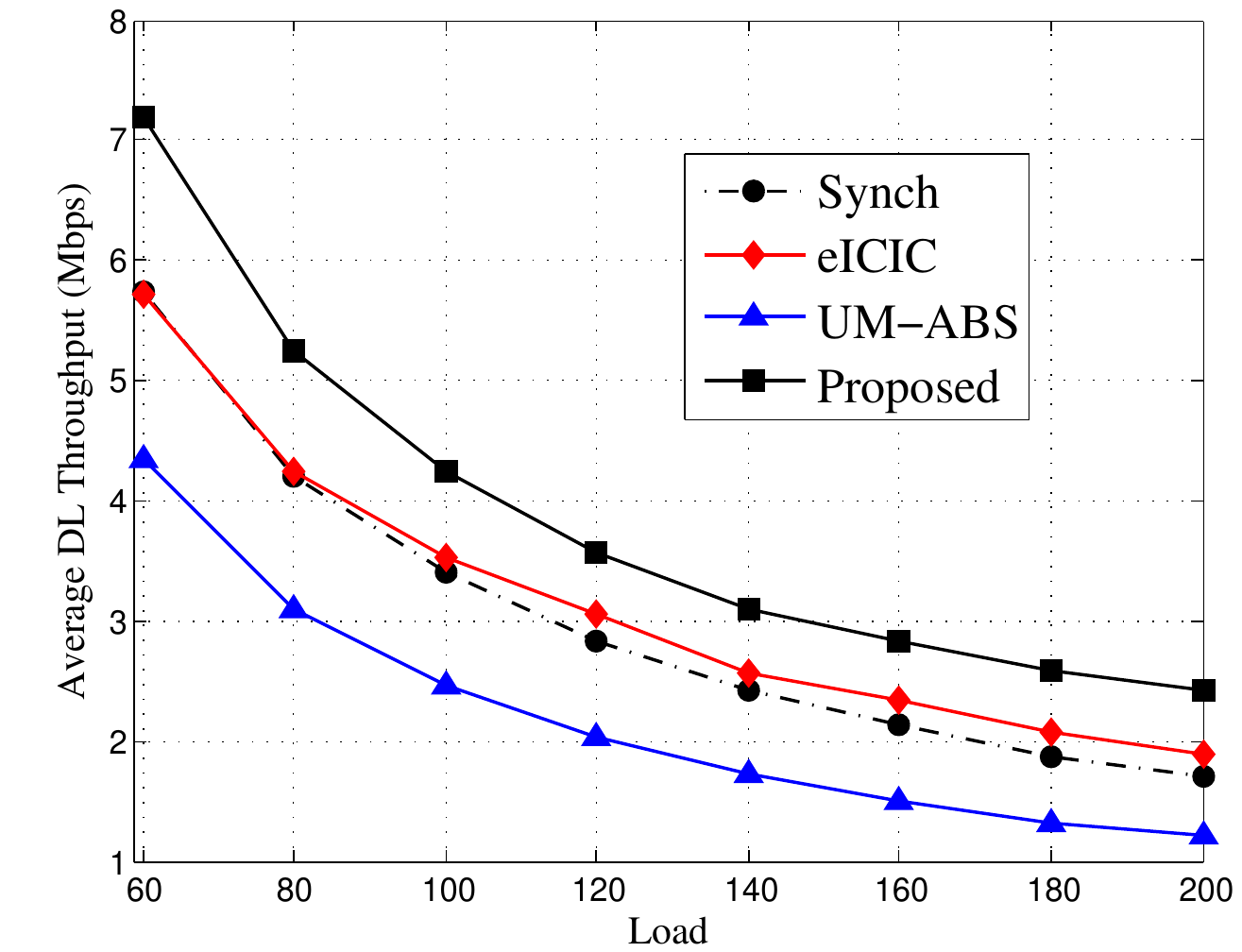}
\end{minipage}
}
\subfigure[Performance of UL throughput under different network loads.]{
\begin{minipage}[b]{0.4\textwidth}
\includegraphics[width=1\textwidth]{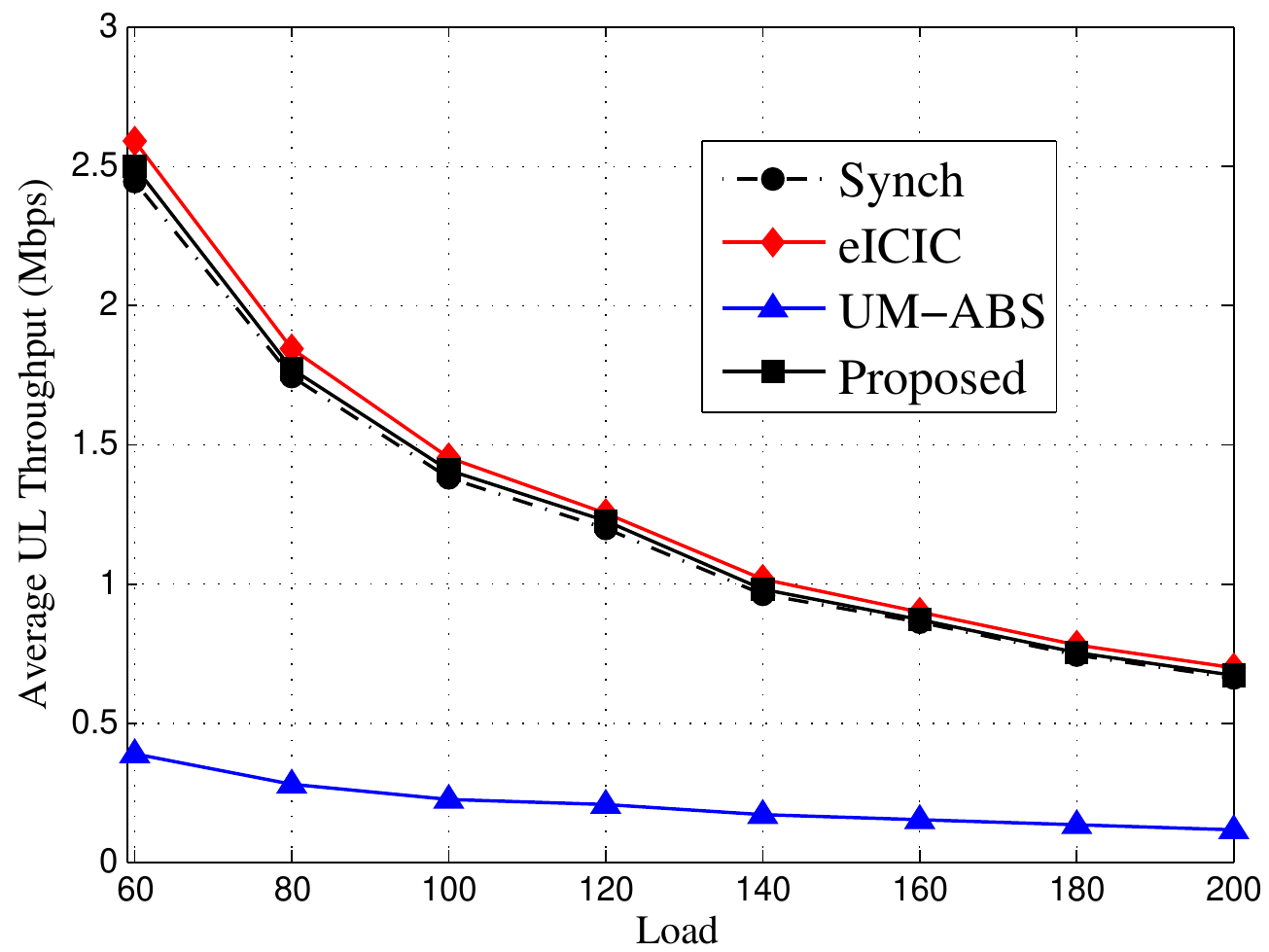}
\end{minipage}
}
\caption{Throughput performance of UEs under different network loads when only BE traffic flows exist.}\label{fig2}
\end{figure}

In Fig. \ref{fig2}, we evaluate network performance in terms of the average DL and UL throughput when network load varies. In our simulations, the network load is the number of UEs located in the HetNet. In this simulation, only BE traffic flows are generated. The average performance of UEs will degrade as the network load increases. However, from Fig. \ref{fig2}(a), it can be seen that in DL direction, the proposed mechanism is superior to other mechanisms, no matter under what kinds of load. 
Besides, from Fig. \ref{fig2}(b), we can see that our proposed mechanism, the eICIC mechanism and the synchronous mechanism will result in similar UL performance, which is also independent of the network load. In addition, the UM-ABS mechanism performs the worst in both directions. This is due to the fact that the interference from SBSs' DL to MBS's UL will also severely degrade the system performance. Such interference does not exist in the other mechanisms except the UM-ABS mechanism.

\begin{figure}[t]
\centering
\subfigure[Performance of DL traffic flows under different network loads.]{
\begin{minipage}[b]{0.4\textwidth}
\includegraphics[width=1\textwidth]{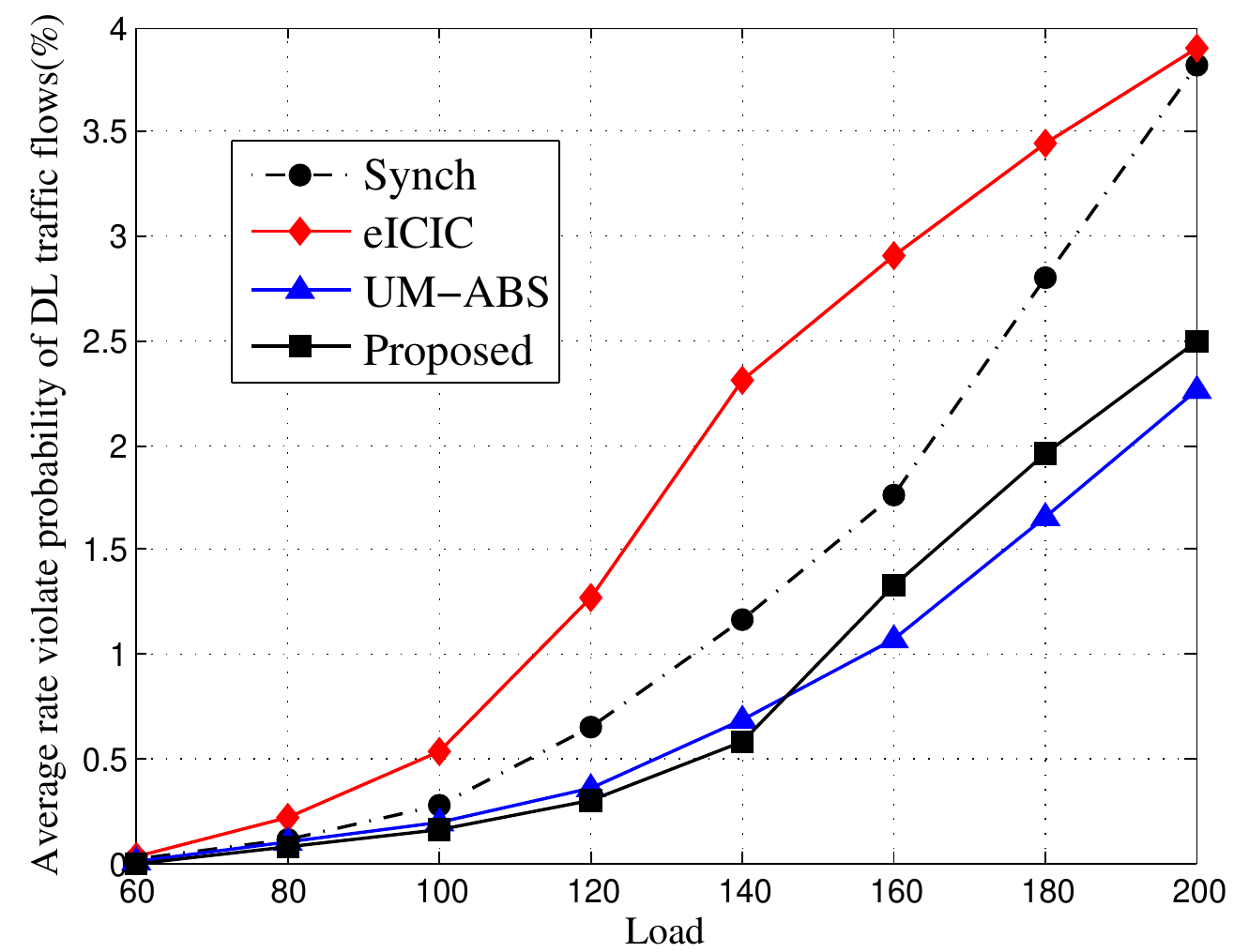}
\end{minipage}
}
\subfigure[Performance of UL traffic flows under different network loads.]{
\begin{minipage}[b]{0.4\textwidth}
\includegraphics[width=1\textwidth]{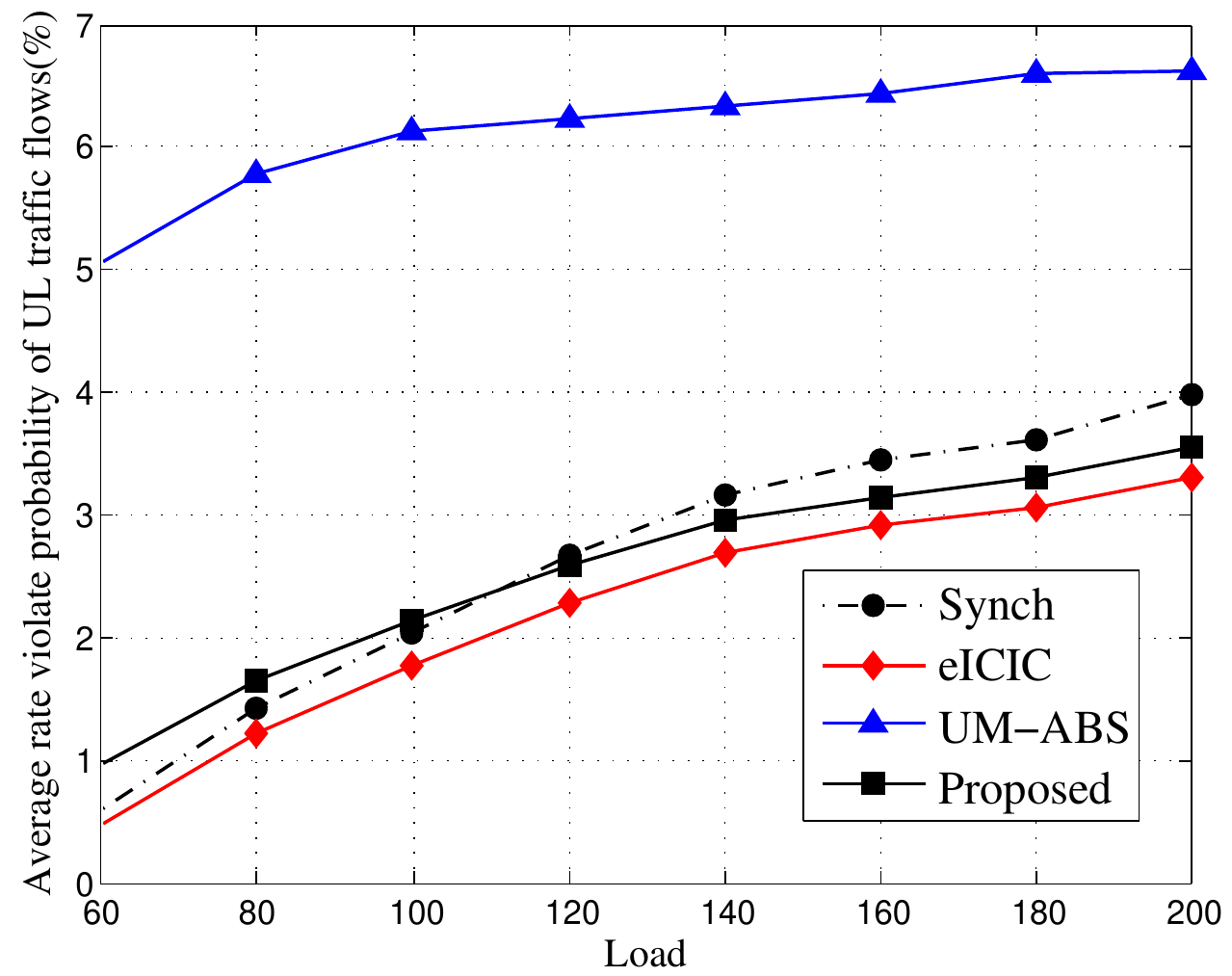}
\end{minipage}
}
\caption{Rate violate probability of soft QoS traffic flows.}\label{fig3}
\end{figure}

Fig. \ref{fig3} depicts network performance when only soft QoS traffic flows are generated. Each soft QoS traffic flow has an intrinsic rate requirement. In this simulation, we define the rate violate probability of traffic flows as the ratio of traffic flows whose intrinsic rate requirements cannot be satisfied. The average rate violate probabilities of traffic flows in both directions increase with the growth of the network load. From Fig. \ref{fig3}(a), it can be seen that the performance of our proposed LP-ABS mechanism is close to that of the UM-ABS mechanism, and is superior to other mechanisms in the DL direction. As shown in From Fig. \ref{fig3}(b), our proposed mechanism, the eICIC mechanism and the synchronous mechanism will result in similar UL performance, which is far better than that of the UM-ABS mechanism. The superiority of our proposed mechanism is due to that severe MBS-to-SBS interference is alleviated and enough resources are reserved for DL and UL transmissions. It should be noted that there exist significant performance differences in the DL and UL directions for the UM-ABS mechanism. This phenomenon can be explained as follows. In the UM-ABS mechanism, DL traffic flows have higher priority to be served due to higher spectral efficiencies, and there is no subframes are reserved for UL transmissions in each SBS.

\begin{figure}[t]
  \centering
  \includegraphics[width=0.4\textwidth]{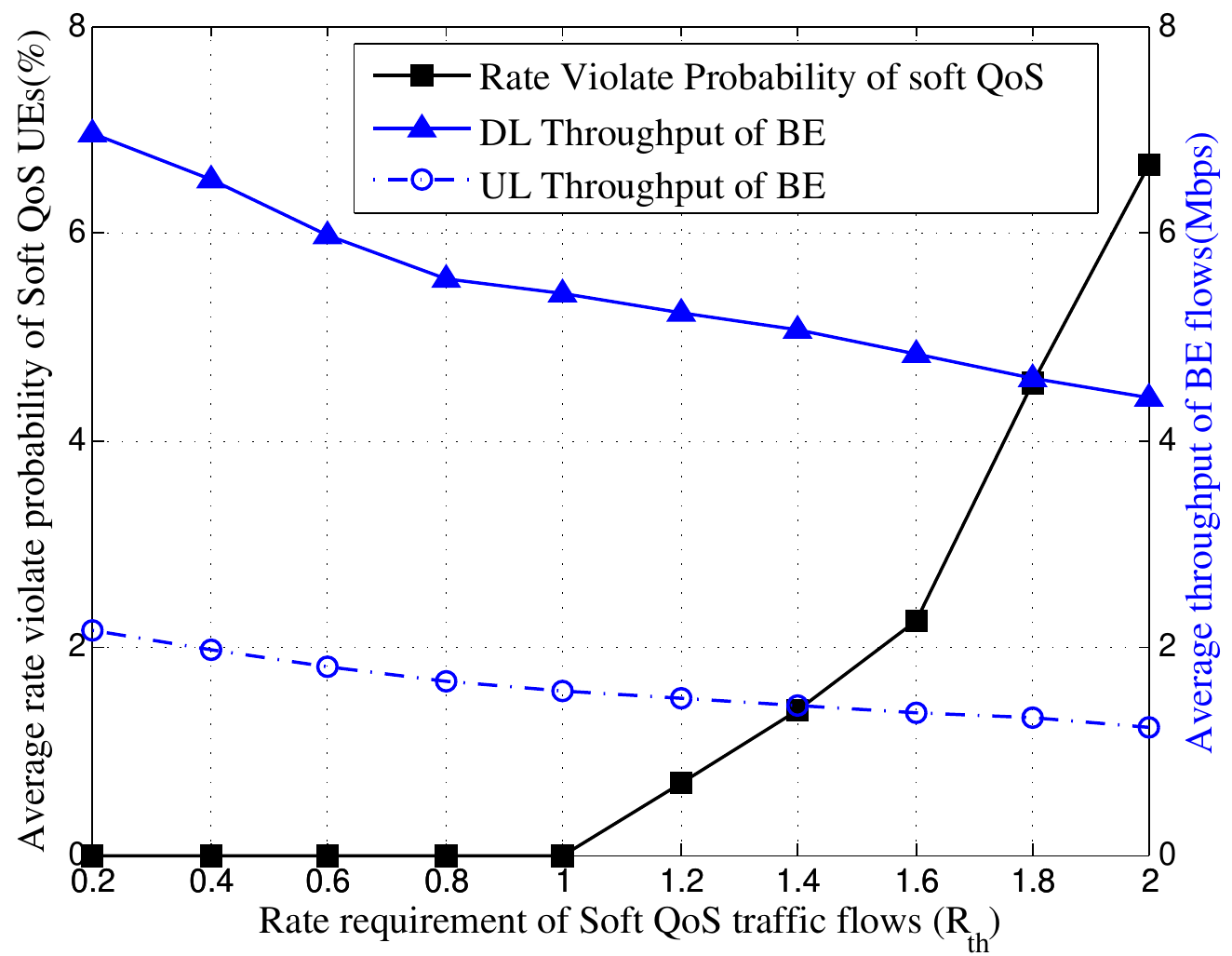}\\
  \caption{Performance of mixed traffic flows.}\label{fig4}
\end{figure}

In Fig. \ref{fig4}, we evaluate the performance of different kinds of traffic flows when mixed traffic flows exist. In this simulation, a total of $120$ UEs are distributed in the network. Half of the UEs generate soft QoS traffic flows in both DL and UL directions, while the other UEs generate BE traffic flows in both link directions. When the rate requirement of sort QoS traffic flows, $R_{th}$, increases from $0.2$ Mbps to $1$ Mbps, the average throughput of BE flows decreases quite slowly, especially in the UL direction. However, the average rate violate probability of soft QoS flows keeps almost zero as $R_{th}$ increases in the above range. This is due to the fact that the proposed resource allocation algorithm gives higher priority to soft QoS traffic flows. Hence, when applying the proposed resource allocation scheme, the BE traffic flows will sacrifice their throughput to guarantee the QoS requirements of soft QoS traffic flows. When $R_{th}$ is large enough, performance of soft QoS traffic flows also degrades because of the resource budget is limited.

\section{CONCLUSIONS}
In this paper, we introduced LP-ABS to alleviate the inherent MBS-to-SBS interference in the dynamic TDD HetNets. With this new approach, the spectral efficiency has been improved and additional flexibility can be obtained to combat traffic fluctuations in small cells. To tackle the resource allocation problem under mixed traffic, we propose a two-step strategy. At BS-level, we define a transmission protocol and divide each transmission cycle into a normal DL/UL period and an LP-ABS period by solving a network capacity maximization problem. At UE-level, we formulate the resource allocation as an NUM problem and solve it with an efficient iterative algorithm. The results from simulations show that the proposed algorithm can improve network throughput as well as QoS satisfaction levels for UEs.

As future work, we will consider dynamic TDD HetNets with mobile UEs. In this case, association and resource allocation should be jointly optimized.


\end{document}